\documentclass[a4paper,pre,twocolumn,amsmath,amssymb,unsortedaddress,showpacs]{revtex4-1}
\usepackage{graphicx}  
\usepackage{dcolumn}
\usepackage{bm}
\newcommand{\be}{\begin{equation}}
\newcommand{\ee}{\end{equation}}

\begin{document}

\title{Time-periodic solitons in a
damped-driven nonlinear Schr\"odinger equation}

 \author{I. V.  Barashenkov}
\email{Email: igor.barashenkov@gmail.com}
\affiliation{Department of  Mathematics,
University of Cape
Town, Rondebosch 7701;}
 \affiliation{National Institute for Theoretical Physics, Stellenbosch, South Africa}
  \affiliation{Joint Institute for Nuclear Research, Dubna, 
 Russia}
 
\author{E. V.  Zemlyanaya}
\email{Email: elena@jinr.ru}
\affiliation{Joint Institute for Nuclear Research, Dubna, 141980 Russia}

\author{T. C. van Heerden}
 \email{Email: tvanheerden@gmail.com}
\affiliation{Department of  Mathematics, University of Cape Town}
 \affiliation{National Institute for Theoretical Physics, Stellenbosch, South Africa}
\altaffiliation{Current address: Department of Applied Mathematics and Theoretical Physics,
Cambridge University, Silver Street, Cambridge CB3 9EW, UK
}

\date{\today}

%
%

\begin{abstract}
Time-periodic solitons of the 
parametrically driven damped
nonlinear Schr\"odinger equation are obtained  as solutions of
the boundary-value problem on a  two-dimensional spatiotemporal domain. 
We follow the transformation of the periodic solitons
as the strength of the driver is varied. 
The resulting bifurcation diagrams provide a natural explanation for the overall form and
details of the attractor chart compiled previously 
via direct numerical simulations.
In particular, the  diagrams
confirm the occurrence of
the period-doubling transition to temporal chaos for small  values
of dissipation and the absence
of such  transitions for larger dampings.
This difference in the soliton's response to the increasing driving strength 
can be traced to the difference in the radiation frequencies in the two cases.
Finally, we relate the soliton's temporal chaos to the homoclinic bifurcation.
\end{abstract}


\maketitle

%
\section{Introduction}
\label{Intro}
 The application of a resonant driving force is an efficient way of compensating dissipative losses in a soliton-bearing system.
If the dissipation coefficient and driving strength are weak,
and the driving frequency is just below the phonon band,
the amplitude of the arising oscillating soliton is 
governed by the nonlinear
Schr\"odinger equation with damping and driving terms. 

The damped-driven nonlinear
Schr\"odinger equations exhibit localised solutions with a variety 
of temporal behaviours, from stationary to periodic and chaotic.
There is a whole range of analytical and numerical approaches to the study of
stationary and steadily travelling solitary waves.
As for the solitons with nontrivial time dependence, such as periodic,
the direct numerical simulation has remained an
exclusive means of obtaining these solutions and classifying their 
stability.

 The shortcoming of this method is that
 simulations capture only {\it stable\/}
 regimes.  
  This means that  the actual mechanisms
 and details of the
transformations of solitons (which are bifurcations involving both stable and
unstable solutions) remain inaccessible. Neither can simulations
be used to identify coexisting attractors in cases of bi-
or multistability.

In this paper we pursue a different approach to the analysis of these
hidden mechanisms. Instead of direct numerical simulations, the
time-periodic solitons are determined as solutions of a
boundary-value problem formulated on a two-dimensional 
spatio-temporal domain.
The advantage of this approach is that it is potentially capable of 
furnishing {\it all\/} solutions --- 
all stable and all unstable.

The particular equation that we are concerned with  here, is the
parametrically driven damped nonlinear Schr\"odinger,
\be 
i \psi_t + \psi_{xx} + 2 |\psi|^2 \psi - \psi = h \psi^* - i
\gamma \psi. 
\label{NLS} 
\ee
In Eq.\eqref{NLS},  $\gamma>0$ is the damping
coefficient and $h$ the amplitude of the parametric driver, which
can also be assumed positive. Equation \eqref{NLS} was used to model a
variety of resonant phenomena in nonlinear dispersive media, including the
nonlinear Faraday resonance in a vertically oscillating water
trough \cite{Miles,LS,Faraday},   formation of oscillons  in granular materials
and suspensions \cite{oscillons},
 synchronisation in parametrically excited pendula arrays
\cite{pendula,Clerc1},
phase-sensitive amplification of light pulses in optical fibers
\cite{optics} and propagation of magnetization waves
in an easy-plane ferromagnet placed in a microwave field \cite{BBK,Clerc1,Clerc2}.
The same equation governs the
amplitude of  breathers in
a variety of systems reducible to  the 
parametrically driven 
damped  sine-Gordon \cite{BBK}
and the $\phi^4$ \cite{JA} equation.
(For more contexts, see \cite{BZ3}.)

The studies of localised states in these mechanical, 
hydrodynamical, magnetic and optical systems have been
 focussing on  structures oscillating {\it periodically}, i.e., without incommensurable frequencies
 in their spectrum.
The {\it quasiperiodically\/} oscillating states
(e.g. states whose fundamental oscillation is modulated by 
smaller frequencies)  have been set aside as complex and atypical. 
 The periodically oscillating  states in the physical systems reducible 
 to equation \eqref{NLS} are described by the stationary  solutions of this equation.
 Accordingly, all previous  analyses of localised solutions of Eq.\eqref{NLS} have been
 confined to  its stationary solitons.
 One of the conclusions of the present project, however,  will be
 that the time-periodic solitons of Eq.\eqref{NLS}
 populate a significant part of its attractor chart. Therefore, the corresponding 
 {\it quasiperiodic\/} localised states should  play 
 a much bigger role in all physical applications of Eq.\eqref{NLS} than it was assumed so far.

The aim of the present work is to follow the transformations
of temporally periodic solitons of equation \eqref{NLS}  as its
parameters are varied, identify the arising bifurcations and
eventually verify and explain the attractor chart for this equation which was
compiled using direct numerical simulations in Ref.\cite{Bondila}.
In the second part of this project \cite{BZ3}
we will complement this one-soliton chart with a 
chart of two-soliton  attractors.

The paper is organised as follows. Section \ref{BG}  (which continues this introduction)
  contains background information on
stationary solitons and their transformations as the parameters of the equation are varied.
It is these transformations that we will be verifying and studying further in the subsequent
sections.

Next, sections \ref{P} and \ref{S} present mathematical techniques we employ in this project:
in section \ref{P} we outline our method of obtaining periodic solitons whereas
section \ref{S} describes our approach to the analysis of their stability.
Section \ref{Rad} introduces a theoretical  framework for the treatment of radiation from 
the oscillating soliton. 

The central results of this study are obtained using numerical methods; these are
reported in section \ref{Num1}.  Here we
present  bifurcation diagrams for the time-periodic free-standing soliton
in various damping regimes. 

The paper is concluded by section \ref{DC} where the results of the 
numerical study are discussed and interpreted.

\section{Stationary and oscillatory solitons:  the background}
\label{BG}
 
Localised stationary or periodic solutions of Eq.\eqref{NLS} 
exist only if $h> \gamma$. 
When $h> h_{\rm cont}(\gamma)$, 
where
\begin{equation*}
h_{\rm cont} = \sqrt{1+ \gamma^2},
\end{equation*}
any localised solution is unstable to continuous-spectrum
perturbations. The evolution of this
instability  leads to spatiotemporal chaos.

Two stationary soliton solutions of Eq.\eqref{NLS} are well known.
One soliton (denoted $\psi_-$ in what follows) exists in the parameter range 
$\gamma \leq h \leq h_{\rm cont}(\gamma)$ and has the form 
\begin{subequations}
\label{solmin}
\begin{equation}
\psi_-(x)= A_-  \exp (- i \theta_-) \,  {\rm sech\/} (A_- x), \label{solmin_a}
\end{equation}
where
\begin{equation}
 A_- = \sqrt{1- \sqrt{h^2-\gamma^2}}, \quad
 \theta_- = \frac{\pi}{2} - \frac12 \arcsin \frac{\gamma}{h}.
\label{solmin_b}
\end{equation}
\end{subequations}
This solution is unstable for all $h$ and $\gamma$ \cite{LS,BBK}.
 We are mentioning this unstable object here because
it will reappear below as a constituent in 
stationary multisoliton bound states. We will also be recalling
this soliton when interpreting complex temporal behaviour of 
time-periodic solutions.

\begin{figure}
\includegraphics[width =\linewidth]{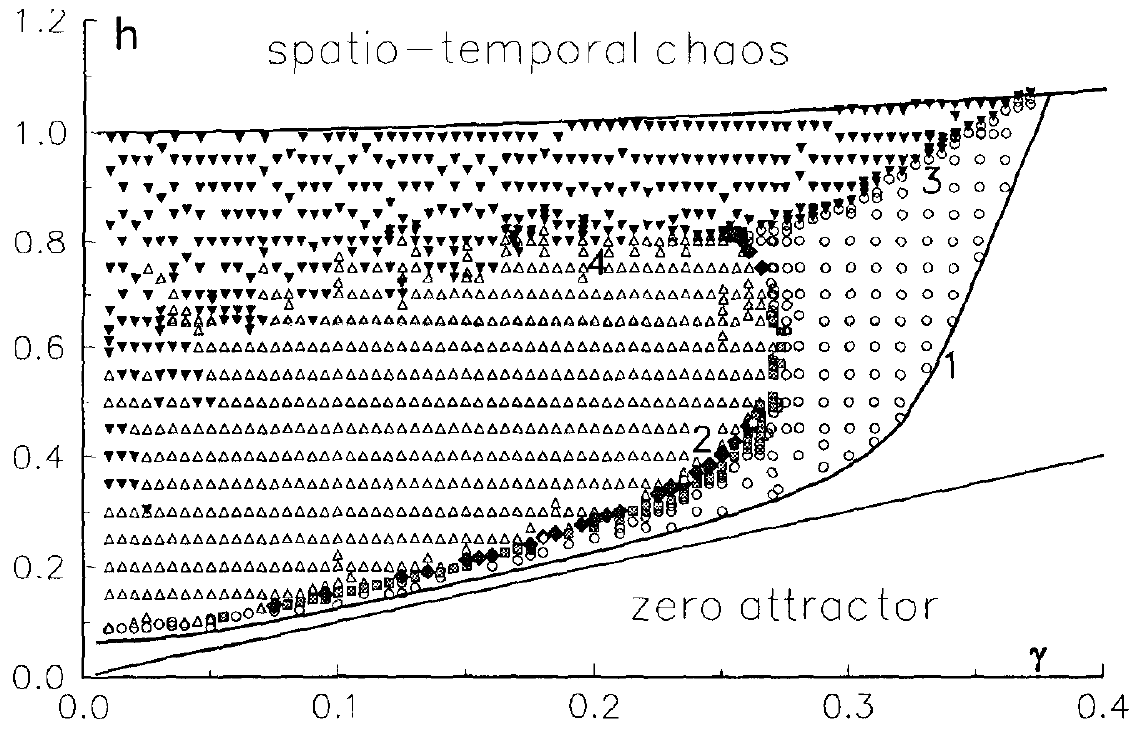}
\caption{\label{chart_from_PhysicaD}The single-soliton
attractor chart for equation \eqref{NLS} compiled by direct numerical simulations  \cite{Bondila}.
Stationary solitons serve as attractors in the blank area above the $h=\gamma$ line.
The area where only the trivial attractor is available is marked by the 
empty triangles.  Empty circles indicate stable periodic solitons; 
the black triangles label spatio-temporal chaos.
Other symbols mark stable higher-periodic and 
temporally chaotic solitons.
For details see \cite{Bondila}. 
} 
\end{figure}

The other stationary soliton exists for all
$h \geq \gamma$; we denote it $\psi_+$:
\begin{subequations}
\label{solplus}
\begin{equation}
\psi_+(x)= A_+ \exp(- i \theta_+) \, {\rm sech\/} (A_+ x), 
\label{solplus_a}
\end{equation}
where
\begin{equation}
 A_+= \sqrt{1+\sqrt{h^2-\gamma^2}}, \quad
 \theta_+ = \frac12 \arcsin \frac{\gamma}{h}.
\label{solplus_b}
\end{equation}
\end{subequations}
The stability properties of this soliton depend on $\gamma$ and $h$ \cite{BBK}.
When $\gamma>0.356$, the $\psi_+$ soliton is stable for
all $h$ in the range $\gamma < h< h_{\rm cont}(\gamma)$. 
When $\gamma< 0.356$, on the other hand, the 
 soliton \eqref{solplus} is only stable for $\gamma< h < h_{\rm
Hopf}(\gamma)$, where the value $h_{\rm Hopf}(\gamma)$ lies
between $\gamma$ and $h_{\rm cont}(\gamma)$
(see the curve labelled $1$ in Fig.\ref{chart_from_PhysicaD}).
 As we increase $h$ past $h_{\rm Hopf}(\gamma)$
 keeping
$\gamma$ fixed, the stationary soliton loses its stability to a
time-periodic soliton \cite{BBK,ABP}. The transformation scenario
arising as $h$ is increased further depends on the choice of the (fixed) value of
$\gamma$. 

The numerical simulations
\cite{Bondila} (also \cite{FLS}) indicate that  
for $\gamma$ smaller than approximately $0.25$, the periodic
soliton follows a period-doubling route to temporal chaos. In 
a wide region of
$h$ values
above the chaotic domain, the equation does not support any stable
spatially-localised solutions. In this ``desert" region,
 the only attractor found in direct numerical
simulations was the trivial one, $\psi=0$. Finally, for even
larger values of $h$, the unstable soliton seeds the spatio-temporal
chaos \cite{Bondila} (see also \cite{FLS}). 

As $h$ is increased for the fixed $\gamma$ greater than 
approximately $0.275$, 
the soliton follows a different
transformation scenario. Here, the period-doubling cascade does
not arise and the soliton death does not occur. The periodic
soliton remains stable until it yields directly to a
spatio-temporal chaotic state \cite{Bondila}.

In a short intermediate range of $\gamma$-values,
$0.25 < \gamma <0.275$, we have a combination of the 
above two scenarios. The increase of $h$ for the fixed
$\gamma$ results in 
the period-doubling of the soliton, culminating 
in the
temporal-chaotic regime,
which is followed 
by the  soliton death. As we continue
to raise $h$, an inverse sequence of bifurcations is observed which
brings the stable single-periodic soliton back. On
further increase of $h$, it loses its stability to a
spatio-temporal chaotic state \cite{Bondila}.

The attractors arising for various 
$h$ and $\gamma$ values are illustrated by Fig.\ref{chart_from_PhysicaD}
which we reproduce from Ref.\cite{Bondila}. We should emphasise that 
this attractor chart has been compiled using direct numerical
simulations of equation \eqref{NLS}, with a particular choice of 
initial conditions. (The initial condition was chosen 
in the form of the unstable soliton $\psi_+$ --- in the parameter range where it
is unstable.) It is an open question, therefore,
how robust this chart is. Would it change if simulations started with 
a different initial condition, or if one used a numerical scheme with different 
parameters (such as the spatial interval length, the number of the Fourier modes,
the full time of simulation,  temporal stepsize,
 etc)? In particular, would qualitative 
features of the chart survive, e.g.  the coexistence of two
transformation scenarios,  appearance of the desert region, and
the peculiar shape of the periodic-attractor domain?
One of the aims of the present work is to answer these questions.


\section{Periodic solitons as solutions of a boundary-value problem in 2D}
\label{P}

Instead of solving equation \eqref{NLS} with some initial
condition and determining the resulting attractor by running the
computation for a sufficiently long time, we were searching
 for periodic solutions by solving Eq.\eqref{NLS}
 as a boundary-value problem on a two-dimensional
 domain $(-\infty, \infty) \times (0,T)$. The boundary
 conditions  were set as
\be 
\psi(x,t) \to 0 \quad \text{as} \ x \to \pm \infty, \label{bc}
\ee
and
\be \psi(x,t+T)= \psi(x,t). \label{per}
\ee
The value of $T$ was not available  beforehand;
the period was regarded as an
unknown, together with the solution $\psi(x,t)$.

The periodic solutions were continued (path-followed) in $h$
for the fixed $\gamma$. We employed a predictor-corrector continuation algorithm
\cite{continuation} with a fourth-order Newtonian iteration at
each  $h$. A finite-difference discretization with the stepsize
$\Delta x=0.05$ was used on  the interval $(-L,L)=(-50,50)$.

\section{Stability of periodic solutions}
\label{S}
\subsection{Floquet multipliers}

Let  $\psi_0(x,t)= {\cal
R}(x,t)+ i {\cal I}(x,t)$ be a spatially localised, time-periodic solution. 
Letting
$\psi(x,t)=\psi_0(x,t) + u(x,t)+i v(x,t)$ and linearising
\eqref{NLS} in the small perturbation $u+iv$, we obtain
\be
 J {\bf w}_t=({\cal H}- \gamma J){\bf w},
 \label{linearised}
\ee
where ${\bf w}={\bf w}(x,t)$ is a two-component column-vector
\[
{\bf w}= \left( \begin{array}{c} u \\ v \end{array} \right),
\]
$J$ is a skew-symmetric matrix
 \[
J= \left(
\begin{array}{cc}
0 & -1 \\ 1 & 0
\end{array}
\right),
\]
and
$\cal H$ is a hermitian matrix-differential operator
\begin{widetext}
\be {\cal H}= \left(
\begin{array}{cc}
-\partial_x^2 + 1+h - 6{\cal R}^2 - 2 {\cal I}^2 &
- 4 {\cal R}{\cal I}  \\
    - 4 {\cal R}{\cal I} &
   -\partial_x^2 + 1-h - 2{\cal R}^2 - 6{\cal I}^2
   \end{array} \right).
   \label{Hcal} 
\ee
\end{widetext}

The solution to Eq.\eqref{linearised} with an initial condition
${\bf w}(x,0)$ can be written, formally, as
$
{\bf w}(x,t)= {\cal M}_t {\bf w}(x,0),
$
where the evolution operator ${\cal M}_t$ acts on (vector-)
functions of $x$ but depends, parametrically, on $t$. A
fundamental role is played by eigenvalues of the monodromy
operator ${\cal M}_T$:
\be {\cal M}_T {\bf y}(x)= \mu {\bf y}(x). \label{EV} \ee
Here ${\cal M}_T$ is the evolution operator ${\cal M}_t $
evaluated at $t=T$. The eigenvalues $\mu$  are usually referred
to as the Floquet multipliers and  the exponents $\lambda$, where
$
\mu= e^{\lambda T},
$
 as the Floquet exponents. According to the Floquet theory,
for each $\lambda$ there is a solution ${\bf w}(x,t)$ such that
\be {\bf w}(x,t) = e^{\lambda t} {\bf p}(x,t), 
\label{lp} 
\ee
where ${\bf p}(x,t)$ is periodic with the period 
of the solution $\psi_0$: ${\bf p}(x,t+T)={\bf
p}(x,t)$ for all $t$.

It is useful to establish the relation between the 
evolution operator ${\cal M}_t$ and symplectic maps.
Letting ${\bf w}(x,t)= e^{-\gamma t} {\tilde {\bf w}} (x,t)$, 
Eq.\eqref{linearised} is cast in the form
\be
 J  {\tilde {\bf w}}_t={\cal H}{\tilde {\bf w}}.
 \label{hamsys}
\ee
Since ${\cal H}$ is hermitian, equation \eqref{hamsys} 
is a hamiltonian system (with a quadratic Hamilton
functional). The solution to \eqref{hamsys} with
an initial condition ${\bf w}(x,0)$ is, therefore,
\[
{\tilde {\bf w}}(x,t)  = {\cal S}_t {\bf w}(x,0),
\]
where ${\cal S}_t$ is a symplectic map. Thus
the evolution operator ${\cal M}_t$ and, in particular,
the monodromy operator ${\cal M}_T$, are related to
symplectic maps: ${\cal M}_t = e^{-\gamma t} {\cal S}_t$
and
\be
{\cal M}_T = e^{-\gamma T} {\cal S}_T.
\label{relation}
\ee
We will use this relation below, in order to explain symmetries 
of the set of eigenvalues of the operator ${\cal M}_T$.

\subsection{Spectrum structure}

Stability properties of {\it stationary\/} solutions are determined 
by eigenvalues $\lambda$ of the operator $(J^{-1} {\cal H}-\gamma)$. 
The corresponding Floquet multipliers are given by $e^{\lambda T}$
(where $T$ can be chosen arbitrarily for a stationary solution). 
Therefore the stability eigenvalues $\lambda$ are nothing but the Floquet 
exponents.
Each of the two stationary solitons $\psi_+$ and $\psi_-$ has just 
one zero stability eigenvalue 
(i.e. just one exponent $\lambda=0$). This zero mode originates 
from the translation invariance of equation \eqref{NLS} --- 
this is the only continuous symmetry the damped-driven 
equation has.

At the point of the Hopf bifurcation of  a stationary solution
(a free-standing soliton $\psi_+$ or a stationary bound state of solitons),
two complex-conjugate eigenvalues $\lambda, \lambda^*$ 
 cross
into the ${\rm Re} \lambda >0$ half-plane. Irrespectively of
what one takes for the value of $T$
in this case, the corresponding Floquet multipliers
$\mu=e^{\lambda T}$ and $\mu^*= e^{\lambda^* T}$ cross through the
unit circle on the complex $\mu$-plane.  If we, however, take $T$
to be equal to the period of the periodic solution
 bifurcating off at this point, i.e., let $T= 2 \pi/ {\rm Im} \, \lambda$, the
Floquet multipliers cross through the unit circle exactly at $\mu=1$.
Adding the unit multiplier associated with the translation
invariance, we conclude that   
any stationary solution has 3 unit Floquet multipliers
at the point of its Hopf bifurcation and, accordingly, the detaching
periodic solutions should also have 3 unit multipliers at that point.

As we continue the periodic solution 
$\psi_0(x,t)$ away from the point where
it was born, two of the unit Floquet multipliers persist
(while the third one moves away along the real axis). These two
unit multipliers are associated with the translation invariance
in $x$ and periodicity in time, respectively.
Indeed, if we substitute $\psi_0(x,t)$ back in Eq.\eqref{NLS} and
differentiate the resulting identity with respect to $x$ and $t$,
we will obtain Eq.\eqref{linearised} with ${\bf w}=\partial_x
\boldsymbol{\psi}_0$ and ${\bf w}=\partial_t \boldsymbol{\psi}_0$,
respectively, where the two-component vector
\[
\boldsymbol{\psi}_0= \left( \begin{array}{c} {\cal R}(x,t)\\ {\cal
I}(x,t) \end{array} \right).
\]
This means that equation \eqref{linearised} has two {\it periodic\/}
solutions and the monodromy operator \eqref{EV} has two 
eigenvalues $\mu=1$ with eigenfunctions
${\bf w}=\partial_x \boldsymbol{\psi}_0$ and ${\bf w}=\partial_t
\boldsymbol{\psi}_0$,
 respectively. These two unit eigenvalues
will routinely arise in our stability analysis of periodic
solutions of Eq.\eqref{NLS}.

It is not difficult to show that real eigenvalues of ${\cal M}_T$ will
always arise in pairs whereas the complex eigenvalues will 
always come in quadruplets. Indeed, the monodromy operator 
is related to a symplectic map by the relation
\eqref{relation}. Real eigenvalues of symplectic maps are
known to come in pairs; if $\nu$ is a real eigenvalue
of ${\cal S}_T$, then so is $\nu^{-1}$. Complex eigenvalues 
of ${\cal S}_T$ come in quadruplets; if $\nu$ is such an eigenvalue,
then so are $\nu^{-1}, \nu^*$, and $(\nu^*)^{-1}$ \cite{Arnold}.
The relation \eqref{relation} implies then that if 
$\mu$ is a real eigenvalue of the evolution operator
${\cal M}_T$, then so is ${\hat \mu}$, 
its inverse with respect to the circle of radius $e^{-\gamma T}$:
\[
\mu {\hat \mu} = e^{-2 \gamma T}.
\]
In a similar way, if $\mu$ is a complex eigenvalue of ${\cal M}_T$, 
then $\mu^*$, ${\hat \mu}$,  and ${\hat \mu}^*$ are eigenvalues
as well. 

We will be referring to ${\hat \mu}$ simply
as  the mirror image of $\mu$. 
Out of the two Floquet multipliers $\mu$ and ${\hat \mu}$ only one 
can cross the unit circle whereas its mirror image will always be confined
inside a smaller circle of the radius  $e^{-\gamma T}$.
Consequently, when analysing the motion of multipliers on the 
complex plane and resulting stability changes, we will be 
focussing on $\mu$ with moduli greater than $e^{-\gamma T}$ and 
ignoring those with $|\mu|<e^{-\gamma T}$.

Finally, we discuss the location of the continuous spectrum.
As $|x| \to \infty$, the operator $(J^{-1}{\cal H}-\gamma)$ becomes a matrix differential
operator with constant coefficients whose spectrum is easily
determined. Namely, when $h^2<1$, the spectrum consists of
all $\lambda$ of the form $\lambda=-\gamma \pm ip$, with
$\sqrt{1-h^2} \leq p \leq \infty$. The corresponding Floquet
multipliers fill in a circle of the radius $e^{-\gamma T}$ on the complex
$\mu$-plane. When $h^2>1$, the continuous spectrum of $(J^{-1}{\cal H}-\gamma)$
fills the vertical line ${\rm Re} \, \lambda=-\gamma$
and an interval on the real axis, $-\gamma-\sqrt{h^2-1}< \lambda
< -\gamma + \sqrt{h^2-1}$. The corresponding Floquet multipliers
fill the circle of the radius $e^{-\gamma T}$ and in addition,
an interval on the real axis: $e^{-(\gamma+\sqrt{h^2-1})T}< \mu
< e^{-(\gamma -\sqrt{h^2-1})T}$.

\subsection{Numerical stability analysis: the method}

To find out whether equation \eqref{linearised} admits solutions
of the form \eqref{lp} with $\mbox{Re} \lambda>0$, we expand
$u(x,t)$ and $v(x,t)$ in the Fourier series on the interval
$(-L,L)$.
The bulk of our eigenvalue calculations were done with $N=100$ but
we went up to $N=250$ when the eigenfunctions have shown
variations on a small scale.


\section{Numerical study}
\label{Num1}
\subsection{Strong damping: $\gamma=0.30$ and $\gamma=0.35$}

\begin{figure}[t]
\includegraphics[width = \linewidth]{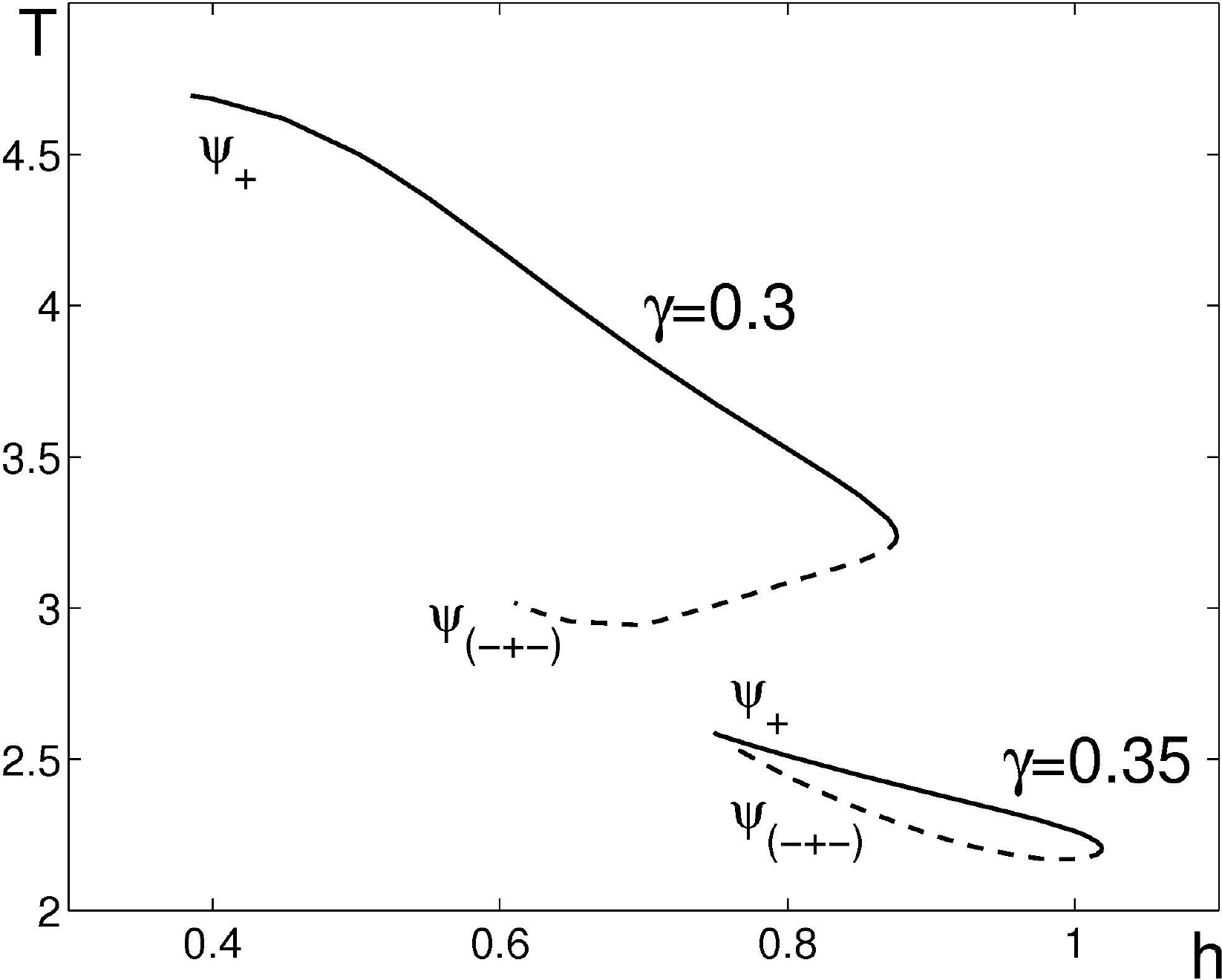}
\caption{The  period  of the periodic
solutions with $\gamma=0.30$ and $\gamma=0.35$. The solid curves show
 stable and
the dashed ones unstable branches. 
\label{ET_03}}
\end{figure}

We explored $\gamma=0.30$ and $\gamma=0.35$
as two representative
sections of the attractor chart in its right-hand part,
where numerical simulations had detected no period-doubling 
bifurcations. The transformation of the solution as it
is continued in $h$ is similar in the two cases; see Fig.\ref{ET_03}.

The top-end point of each of the two
curves  in figure \ref{ET_03}  corresponds
to the stationary single-soliton solution $\psi_+$. 
The underlying value of $h$ equals $0.385$ for $\gamma=0.30$
and $0.7500$  for $\gamma=0.35$.
At these $h$, the  stationary $\psi_+$ soliton undergoes a Hopf
bifurcation and a stable periodic solution  is  born. 
At the starting point
of each curve, the spectrum
of the periodic soliton
includes three unit multipliers $\mu_{1,2,3}=1$.
As $h$ is increased,
the eigenvalue $\mu_3$ moves inside the unit circle
whereas $\mu_{1,2}$ remain at $1$. 
Meanwhile, five real positive eigenvalues $\mu_{4,...,8}$
detach, one after another, from the continuous
spectrum.
One of these ($\mu_8$) later returns to the continuum while the other 
four  eigenvalues move 
towards the unit circle. 
At some point the eigenvalue
$\mu_4$ collides with  $\mu_3$
producing a complex pair which, however, later returns to the positive
real axis.  At the turning point $h_{\rm sn}$, two positive eigenvalues,
$\mu_3$ and $\mu_4$,
cross through the unit circle (almost simultaneously). 
 This is where the 
periodic solution loses its stability.
Numerically, the turning-point value is $h_{\rm sn}=0.8761$
for $\gamma=0.30$ and $h_{\rm sn}=1.0186$
for $\gamma=0.35$.
 On the unstable branch, the spectrum includes two
positive eigenvalues $\mu_{3,4}>1$, two unit eigenvalues $\mu_{1,2}=1$,
and
one positive eigenvalue close to unity ($\mu_5<1$).
In addition, two more positive eigenvalues approach the  
unit circle from inside as we continue away from the turning point.

\begin{center}
\begin{figure}
\includegraphics[height = 1.7in, width = \linewidth]{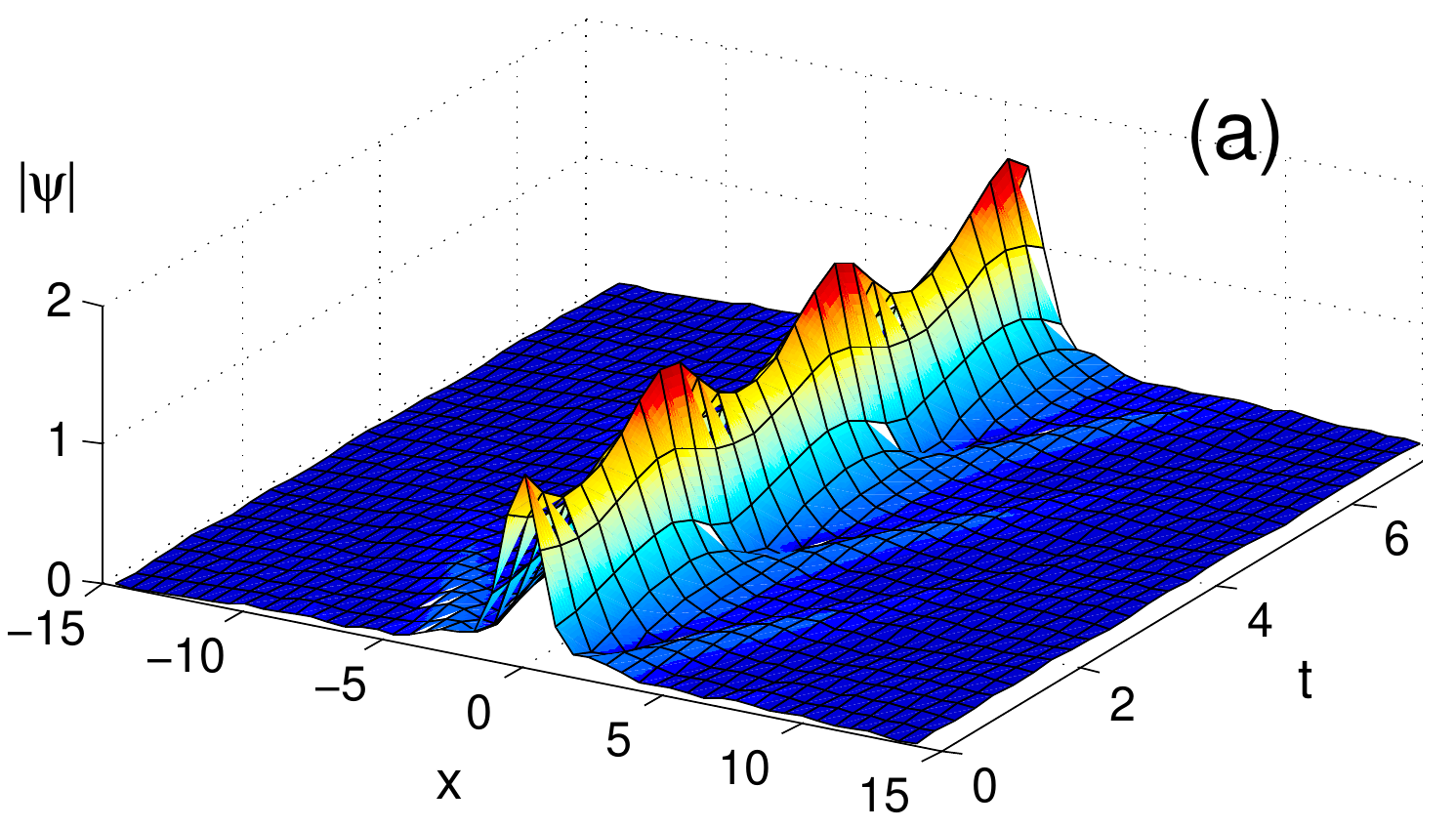}
\includegraphics[height = 1.7in, width = \linewidth]{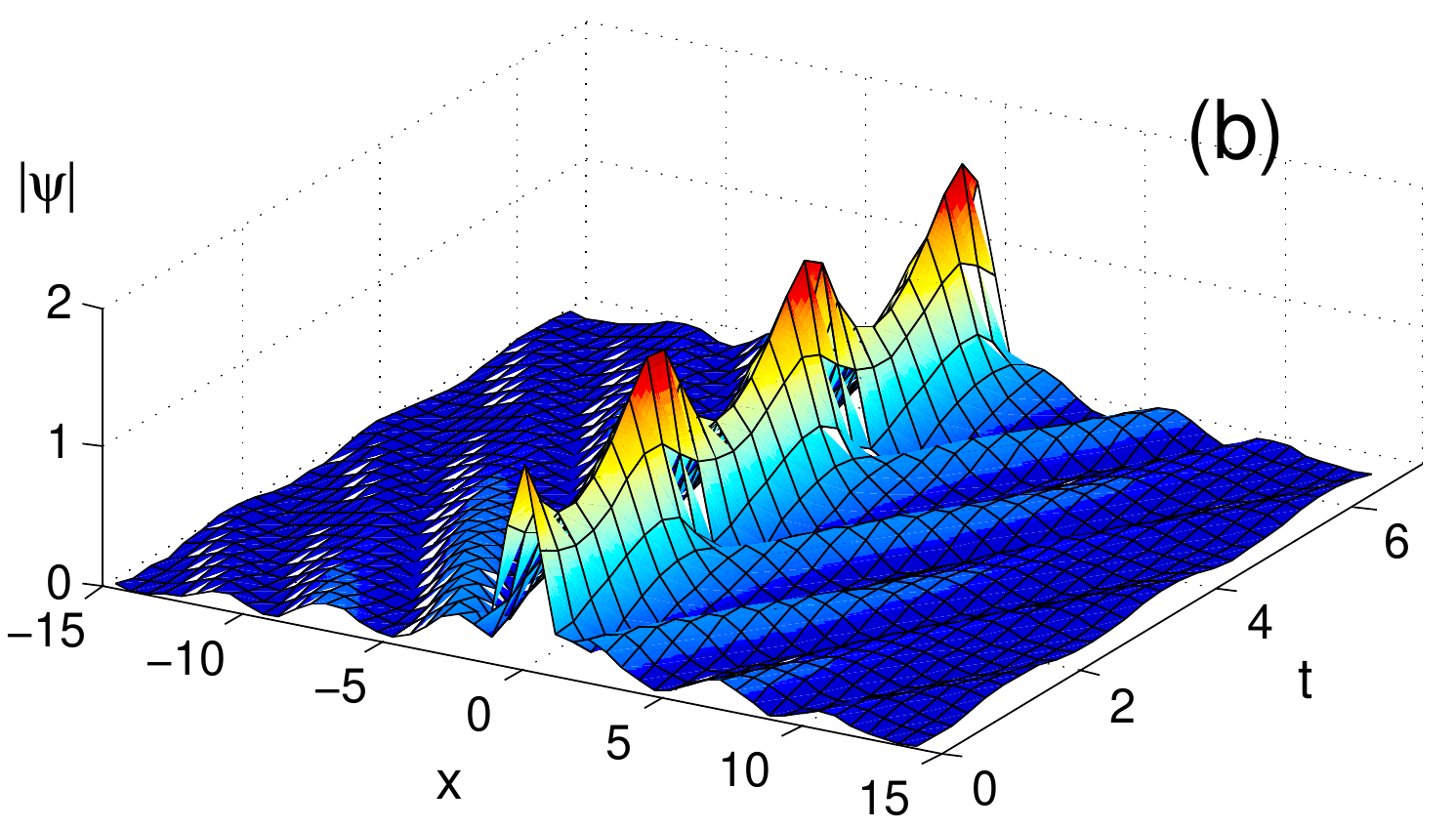}
\includegraphics[height = 1.7in, width = \linewidth]{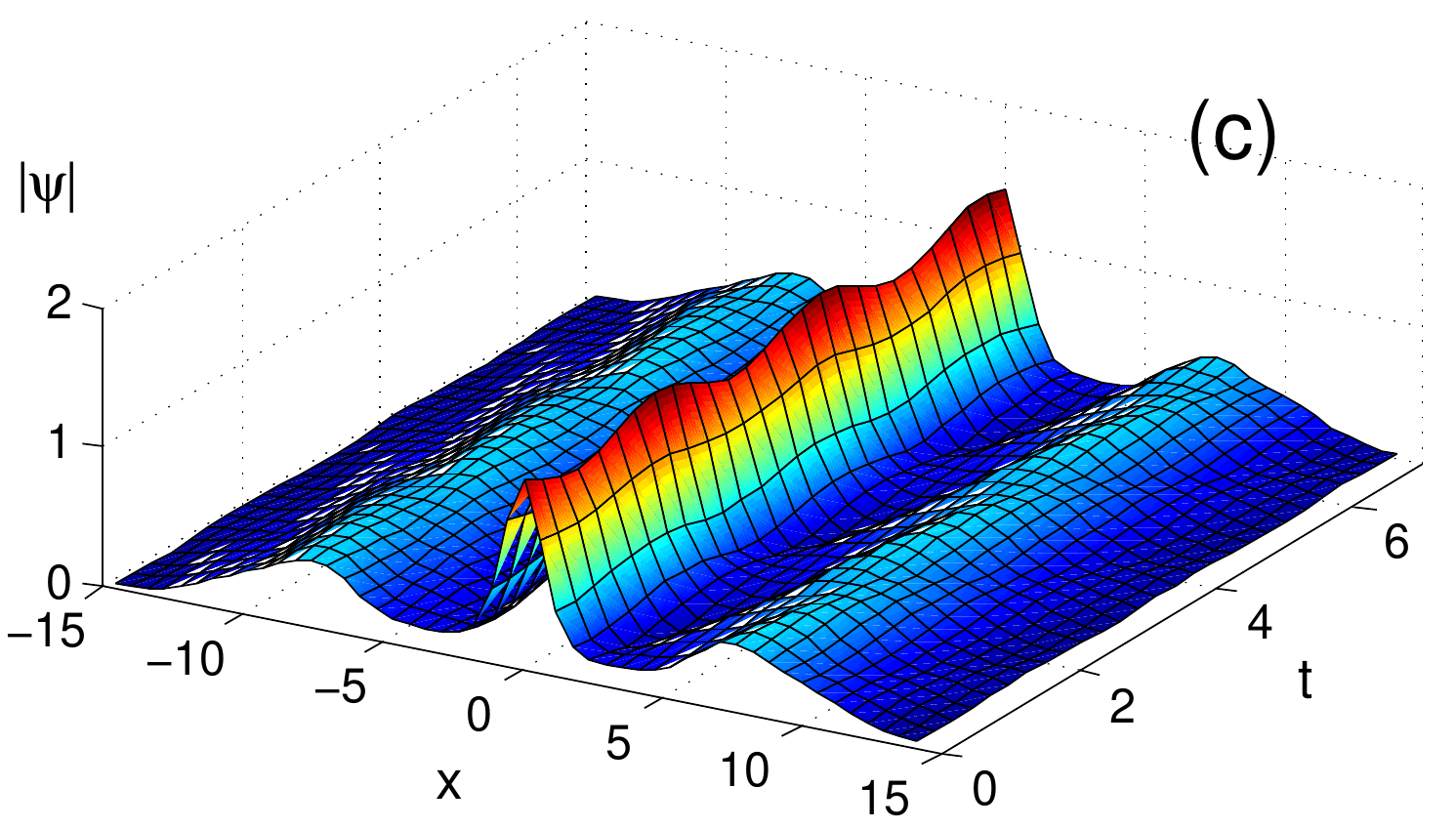}
\caption{\label{g03_3D} (Color online)
The absolute value of the periodic
solution with large $\gamma$. In this plot,
$\gamma=0.35$. ``Motion pictures"  (a) and (c) have been
 taken at the same value of $h$ but correspond to the different branches of
 the diagram Fig.\ref{ET_03}: (a) $h=0..85$, $T=2.45$
 (top branch) and (c) $h=0.85$, $T=2.34$ (bottom branch). 
The evolution (b) corresponds to the turning point: $h=1.0185$, $T=2.21$.
In each case
several periods of oscillation are shown. }
\end{figure}
\end{center} 

It is fitting to note here that the two turning point values,  $h_{\rm sn}=0.8761$
 and $h_{\rm sn}=1.0186$, are in a good agreement with the boundaries of
 the periodic-attractor existence domain established previously.
 Namely, the direct numerical simulation of Eq.\eqref{NLS} 
 gave values close to $0.86$ and $1.01$ for $\gamma=0.30$ and $0.35$,
 respectively \cite{Bondila}.

 The end point of the dashed curve ($h=0.61$ for $\gamma=0.30$ and
 $h=0.760$
 for $\gamma=0.35$)
corresponds to a stationary  complex of solitons.
In the  $\gamma=0.35$ case, this solution has three
separate humps in its real and imaginary part. 
Comparing (the real and imaginary parts of)
each of the three humps to (the real and imaginary parts
of) the free-standing solitons $\psi_+$ and $\psi_-$, 
we conclude that the complex consists of the $\psi_+$ soliton 
in the middle and two solitons $\psi_-$ on its sides;
that is, the stationary complex should be interpreted as 
$\psi_{(-+-)}$.  The spectrum of discrete eigenvalues of this 
stationary complex 
is close to a union of the eigenvalues of the
 soliton $\psi_+$ and those of two solitons $\psi_-$.
Namely, we have three unit eigenvalues $\mu_{1,2,5}=1$
(resulting from the translation modes of two $\psi_-$'s and one
$\psi_+$)
and  two eigenvalues $\mu_{6,7}$ close to unity (contributed by the 
$\psi_+$ soliton which is close to its Hopf bifurcation point). 
In addition, the two
$\psi_-$ solitons contribute two real positive eigenvalues $\mu_{3,4}>1$.

In the case of $\gamma=0.30$, 
the two side peaks in the real part of the end-point solution
are seen to have merged with the central peak (whose
height coincides with the height of the $\psi_+$
soliton in its real part). 
On the other hand, the central peak in the 
imaginary part of the solution is
seen to have disappeared while the two lateral peaks have
the same height as the soliton $\psi_-$ in its imaginary part. 
This indicates that the solution can still be interpreted as 
the $\psi_{(-+-)}$ complex,
albeit a tightly bound one.

Representative solutions  are shown in figure \ref{g03_3D}. 
Near the leftmost point of the curve pertaining to $\gamma=0.35$ in
Fig.\ref{ET_03}, the periodic solution looks like a single soliton
with a periodically oscillating amplitude and width
[Fig.\ref{g03_3D}(a)]. The soliton is emitting radiation waves; however,
the radiation is decaying rapidly as $|x| \to \infty$. As we move further along this curve in
Fig.\ref{ET_03}, the amplitude of oscillations as well as the
intensity of radiation increases [Fig.\ref{g03_3D}(b)]. The shape of the oscillating
solution evolves into a three-hump structure. Near the end point of the
curve, the amplitude of oscillations decreases
[Fig.\ref{g03_3D}(c)] and we arrive at the stationary
three-soliton complex.

Thus, periodic solitons with large $\gamma$ connect 
(in the sense of paths in the parameter space) stationary 
solitons and their complexes, with the connection points provided by the Hopf 
bifurcations of the latter.  It is appropriate to mention a recent 
publication \cite{AW} where a similar organisation of the solution manifold
was reported for the Benjamin-Ono equation.

\subsection{Weak damping: $\gamma=0.1$ and $\gamma=0.2$}

\begin{figure}
\includegraphics[width = \linewidth]{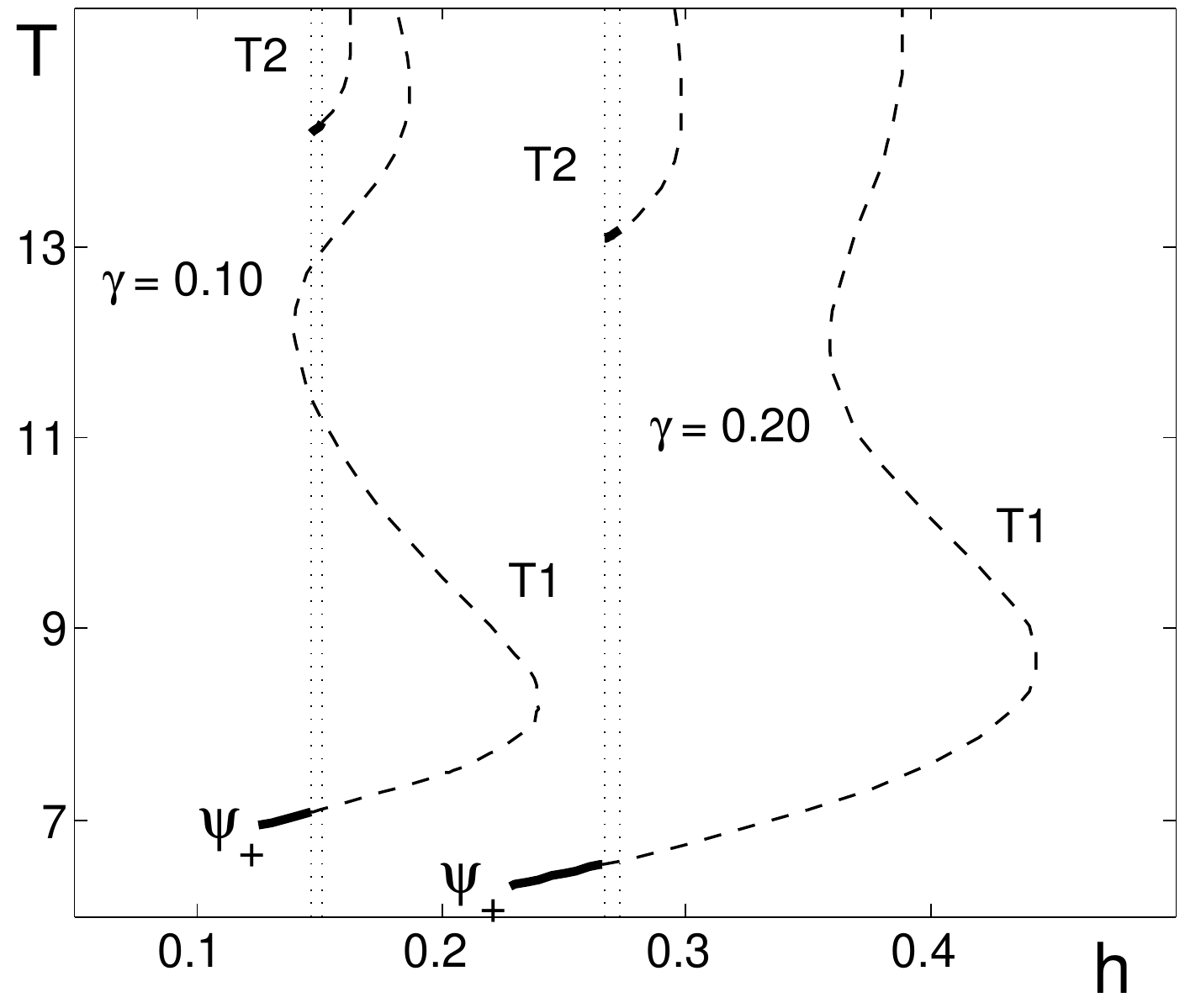}
\caption{The period of the periodic solutions with
 $\gamma = 0.10$ and $\gamma = 0.20$. The solid curves show stable and the dashed one unstable branches.
\label{weak_damping_fig}}
\end{figure}

In  the weak damping regime we explored two representative 
 values of $\gamma$, $\gamma = 0.10$ and $\gamma = 0.20$. The two cases
 exhibit similar bifurcation diagrams (Fig. \ref{weak_damping_fig}) which are, however, very 
 different from the strong-damping diagrams in Fig.\ref{ET_03}.

As before, the starting point (the left-end point)
 of each branch corresponds to the stationary $\psi_{+}$ soliton. 
 At $h = 0.1250$ (for $\gamma = 0.10$) and $h = 0.2275$ (for $\gamma = 0.20$)
 the stationary soliton undergoes a Hopf bifurcation and a periodic solution is born. 
 In Fig.\ref{weak_damping_fig}, this solution is marked $T1$.
 The spectrum of the  $T1$ solution at the starting point of the curve has three unit multipliers, $\mu_{1,2,3} = 1$. 
 As $h$ is increased,  $\mu_{1}$ moves along the real axis inside the unit
 circle and  immerses in the continuous spectrum at
 $\mu=e^{-\gamma T}$. For even a larger $h$, a pair of complex-conjugate multipliers
 with real parts close to  $-e^{-\gamma T}$ detaches from the continuum
 before converging  on the negative real axis. One 
 of the resulting real eigenvalues moves towards $-e^{-\gamma T}$ and rejoins the 
 continuum.
 The other one moves away from the origin and eventually passes
  through $-1$.
This is a signature of the period-doubling bifurcation: the original solution of period $T$ loses its stability and a stable 
solution with a period $2T$ is born. (This happens at $h_{2T} = 0.1463$ and $h_{2T} = 0.2663$ 
for $\gamma = 0.10$ and $\gamma = 0.20$, respectively.)

The spectrum of the newly born $T2$ solution evolves similarly to the spectrum of $T1$. 
It is also born with three discrete multipliers $\mu_{1,2,3}=1$. 
As $h$ increases, $\mu_{1}$ moves along the real axis; immerses in the continuous spectrum;
reappears at $-e^{-\gamma T}$, and eventually 
 passes through $-1$ at some $h=h_{4T}$. (The value $h_{4T}$ equals  $0.1511$  for $\gamma=0.10$
 and $0.2725$  for $\gamma = 0.20$.) 
 At this  point a new stable branch is born, with the period equal to two periods of $T2$.
This new, period-four branch is not shown in Fig. \ref{weak_damping_fig}.

 As for the (unstable)  $T1$ solution, the corresponding $T(h)$ curve is snaking
 up, with the period growing without bound. The solution
approaches a homoclinic orbit, connecting the stationary $\psi_-$ soliton to
itself. In the next subsection, we will discuss this phenomenon in more detail. 
 
 We note that the bifurcation diagrams  of Fig. \ref{weak_damping_fig} 
 are in agreement with direct numerical simulations 
 of Eq.\eqref{NLS} with $\gamma = 0.10$ and $\gamma = 0.20$ \cite{Bondila}
 that revealed the period-doubling transition to temporal chaos in soliton's dynamics.
 It is therefore natural to expect that  our $T4$ branch 
 will also undergo a period-doubling bifurcation after a short interval of stability, 
 and similarly the branches that it births would undergo a sequence of period-doubling bifurcations.
 We did not have computational capacity to verify this using our two-dimensional continuation approach. 
 It is worth mentioning, however, that the values of $h$
 at which higher-periodic and temporally-chaotic attractors were 
 observed in simulations
 ($0.16$ and $0.28$ for $\gamma=0.1$ and $0.2$, respectively \cite{Bondila})
  are close to the $h_{4T}$ given above.  

Finally, we have not been able to perform
an accurate numerical continuation of solutions with $\gamma=0.05$ and
smaller.
In this case the solution  consists of a finite-extent soliton
riding on a (second-harmonic) oscillatory background of a non-negligible amplitude,
which shows only a very slow spatial decay as
$|x| \to \infty$. [Fig.\ref{g03_3D}(b) gives an idea of the shape of the solution in that
case, for most $h$.]
In order to obtain this solution under the boundary conditions
$\psi(\pm L,t)=0$, one has to enlarge the length of the spatial 
side of the domain of computation, $(-L,L) \times (0,T)$. 
This quickly saturates the computational capacity available.

\subsection{Intermediate damping: $\gamma=0.265$.}

Finally, we investigate the transformation of the periodic 
solutions with $\gamma$ lying between the strong- and weak-damping ranges.
As a representative value of such ``borderline" damping, we take 
 $\gamma=0.265$.
According to the numerical simulations,  the {\it direct\/}
period-doubling cascade is followed by an {\it inverse\/} cascade here,
resulting in a peculiar shape of the 
periodicity region on the ($\gamma,h$) plane (Fig.\ref{chart_from_PhysicaD}).
Our aim is to provide an explanation for this phenomenon
on the basis of  the  transformations of the periodic solutions as
$h$  is continuously varied. 
\begin{figure}[t]
\includegraphics[width = \linewidth]{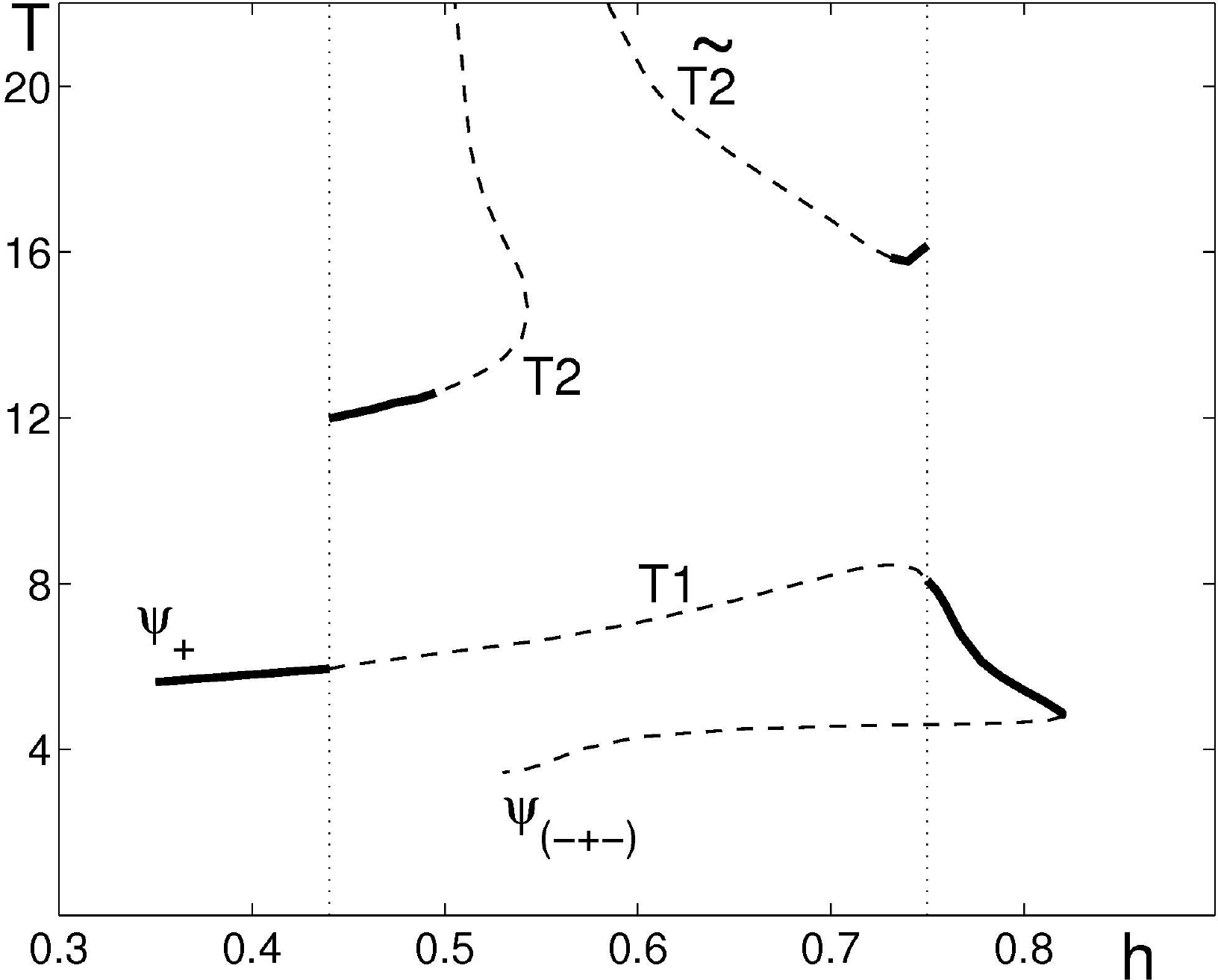}
\caption{\label{Fig_265}
 The  period  of the
periodic solution for $\gamma=0.265$. Solid curves show  stable
and the dashed ones unstable branches.}
\end{figure}

The results of our numerical continuation are summarised in Fig.\ref{Fig_265}.
The bifurcation diagram consists of three branches, the 
second and third of which
arise as a result of the period-doubling bifurcation of the
solution on the first branch. The transformation of the solution
as we move along this period-one branch [the bottom branch in
Fig.\ref{Fig_265}, denoted $T1$] is similar to the transformation
of the solution along the curves shown in Fig.\ref{ET_03}. The
starting point of the curve (its leftmost point) corresponds to
the stationary solution $\psi_+$; the end point corresponds to the
stationary complex $\psi_{(-+-)}$.
  The major difference from the
case of large $\gamma$ ($\gamma=0.3$ and $\gamma=0.35$) occurs 
when a real eigenvalue crosses through
$-1$ as $h$ is increased through $h=0.44$. As $h$ is increased
past $0.44$, this negative eigenvalue continues to grow in
absolute value, reaches a maximum, and then starts to decrease. As
$h$ is increased through $h=0.75$, the negative eigenvalue crosses
through $-1$ once again, this time in the direction of decreasing
modulus. The described behaviour of the negative
eigenvalue corresponds to period-doubling bifurcations at $h=0.44$
and $h=0.75$, where  periodic solutions with double period are
detached.  For larger $h$ the solution is stable --- all the way to
the turning point,
where two real eigenvalues cross out 
through the unit circle.

The spectrum of the first double-periodic solution (detaching at
$h=0.44$ and denoted $T2$ in Fig.\ref{Fig_265}) includes two unit eigenvalues,
$\mu_{1,2}=1$. When $h=0.44$, there is also a negative
eigenvalue $\mu_3=-1$. As $h$ is increased from $0.44$,
 $\mu_3$ moves inside the unit circle, 
 then reverses and crosses through $ -1$
once again (at $h_{4T}=0.495$). At this point, a new periodic solution
$T4$
is born, with the period equal to double the period of the
solution $T2$ --- roughly four times the period of $T1$.
(This ``period-four" solution is not shown in Fig.\ref{Fig_265}.)
The value $h_{4T}=0.495$ coincides with the largest value of $h$
at which higher-periodic attractors were seen in simulations \cite{Bondila}.

\begin{center}
\begin{figure}[t]
\includegraphics[height = 1.7in, width = \linewidth]{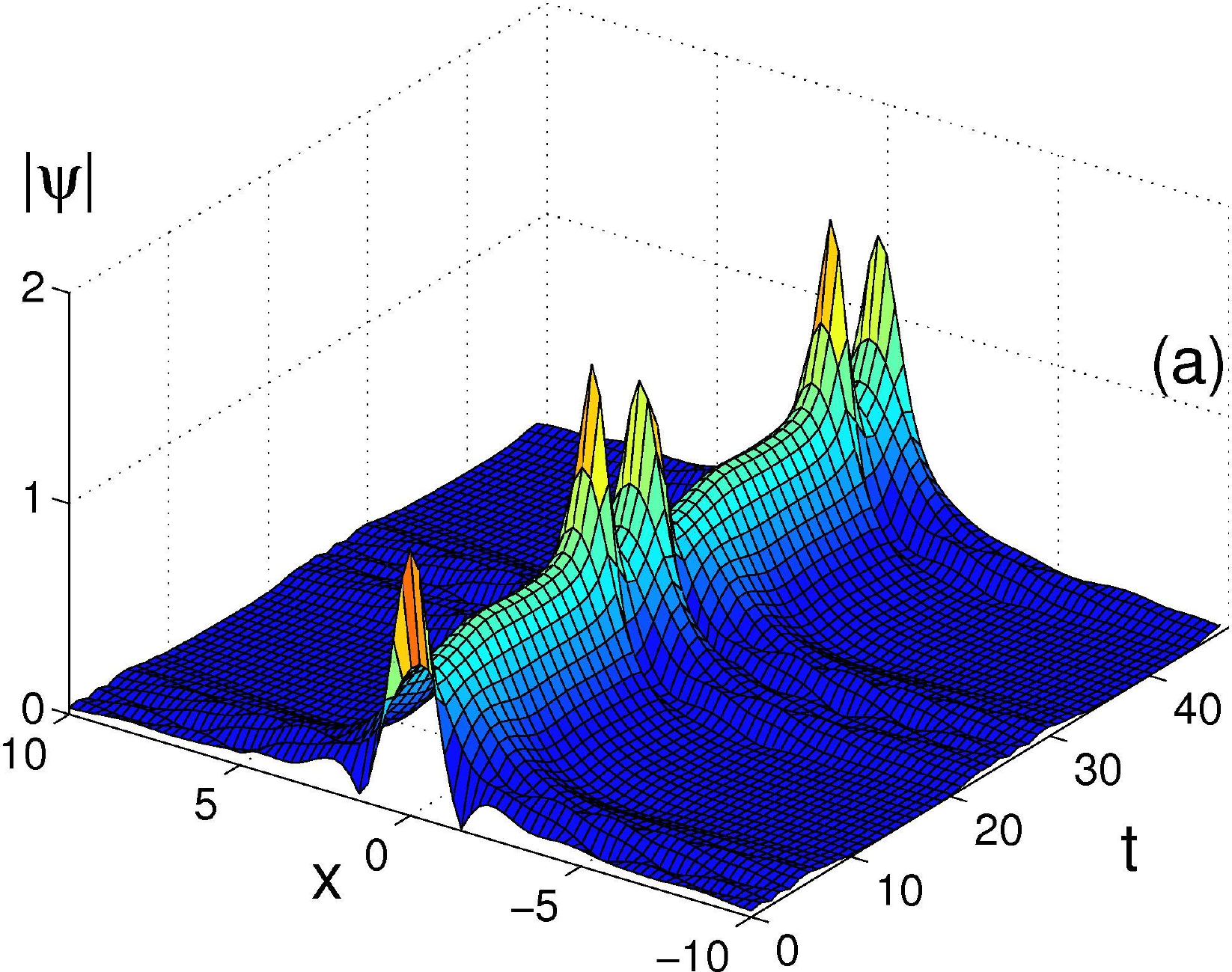}

\vspace*{3mm}
\includegraphics[width = \linewidth]{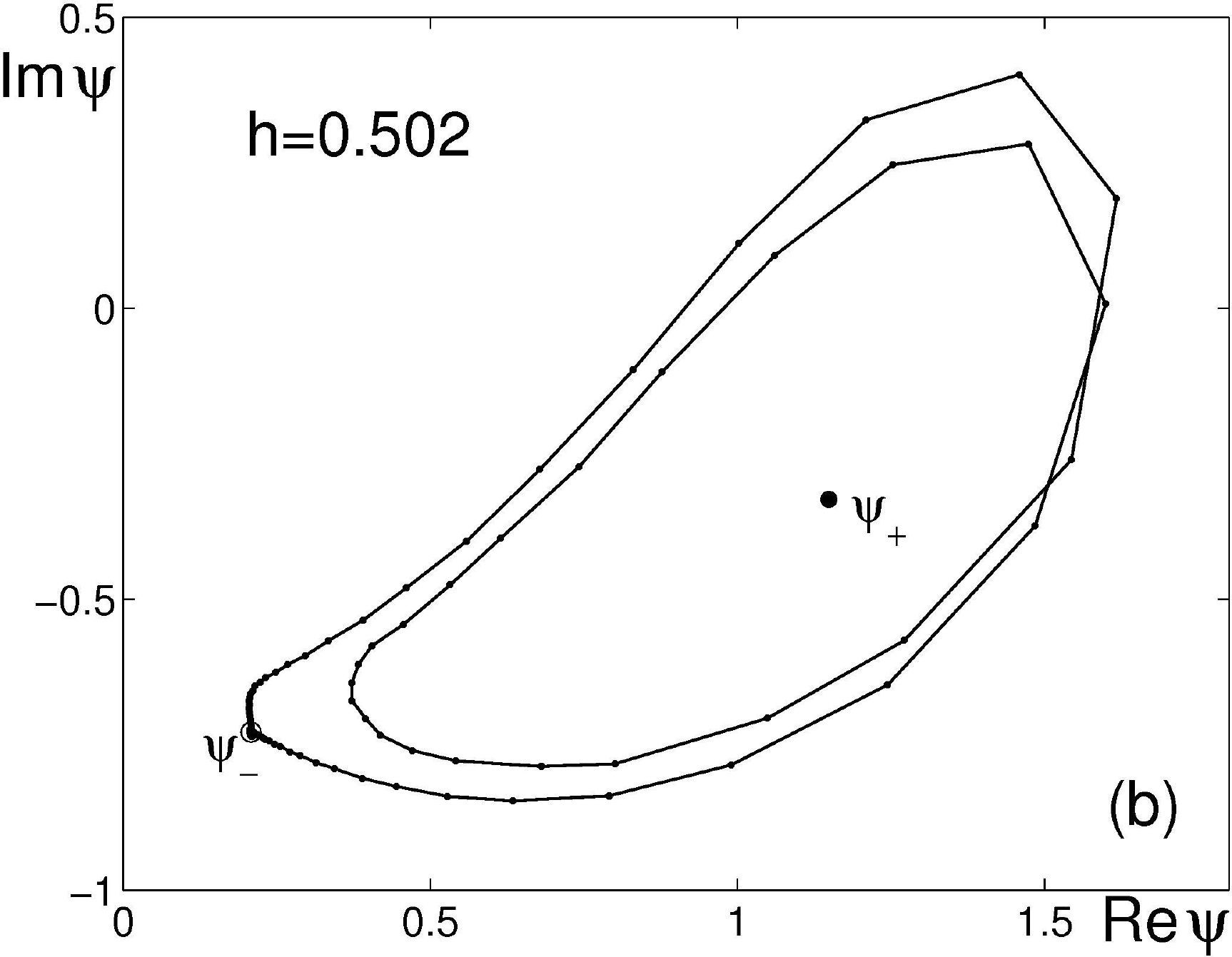}
\caption{\label{Fig_2650} (Color online) 
 (a) The absolute value of the double-periodic solution
 found on the left upper branch in Fig.\ref{Fig_265}. Here $\gamma=0.265$, $h=0.502$ and
 $T=22.985$. A rapid oscillation
 is followed by a long quasistationary
epoch where the solution remains close to the $\psi_-$ soliton. 
(b) The phase portrait of this solution taken at $x=0$. Shown is 
${\rm Im} \, \psi(0,t)$  vs ${\rm Re} \, \psi(0,t)$.
The filled and open circle represent two fixed points: 
the stationary solitons $\psi_+$ and $\psi_-$.
}
\end{figure}
\end{center}

Returning to the $T2$ solution, 
it has two positive eigenvalues $\mu_{4,5}$
in addition to the eigenvalues $\mu_{1,2,3}$. As the $T2$ curve 
in Fig.\ref{Fig_265}  is traced
around the turning point at $h=0.543$, these two eigenvalues
cross, almost simultaneously, through the unit circle 
so that $\mu_{4,5}>1$ on the upper
branch of the curve.
 As we continue further along the upper
branch, the period of the solution grows:
  the solution develops a long
epoch where it remains very close to the stationary soliton $\psi_-$.
Each period now consists of two phases:  the solution performs a
rapid oscillation with its spatio-temporal profile  close to that
of the $T1$ solution, followed by a slow passage
through the bottleneck near the $\psi_-$ soliton
[see Fig.\ref{Fig_2650}].

The second period-2 solution 
(denoted $\widetilde{T2}$ in Fig.\ref{Fig_265}), which  
detaches at $h=0.75$, remains stable for
$0.731\leq h \leq 0.75$. As with the first double-periodic
solution $T2$, the period of $\widetilde{T2}$ grows as we move along the
branch.  The solution changes similarly to
what we have described in the previous paragraph: a rapid
oscillation is followed by a long quasistationary epoch when the
solution is very close to the stationary soliton $\psi_-$.

Thus,  periodic solutions on the $T2$ and $\widetilde{T2}$ branches approach a 
homoclinic orbit: an infinite-period solution which tends to the stationary soliton 
$\psi_-$ as $t \to -\infty$ and $t \to \infty$. The same transformation scenario was detected 
in the weak-damping case ($\gamma<0.25$) where the $T1$ (and, presumably, the $T2$) branch was 
seen to snake up to $T =\infty$ (Fig.\ref{weak_damping_fig}).
At the point $h=h_\infty$ where the period becomes infinite, the periodic solution 
undergoes a homocilinic bifurcation which may serve as the source of chaos
(see section \ref{explosion} below).

\section{Radiation from the oscillating soliton}
\label{Rad}

\subsection{The long-range radiation indicator}

Like their  stationary counterparts,
the oscillatory solitons strike the balance between the
energy fed by the driver and the energy lost to the dissipation.
This can be quantified, for example, by considering the integral
$N= \int_{-R}^R |\psi|^2 dx$. [When the nonlinear Schr\"odinger equation is employed to
describe  planar stationary waveguides, this
integral measures  the total power of light 
captured by the $(-R,R)$ section of the waveguide. 
There are energy-related interpretations of this integral in other  contexts as well.]

Equation \eqref{NLS} yields 
\be
{\dot N}= 2h \int_{-R}^R \left.  |\psi|^2 \sin (2 \theta) dx- 2 \gamma N +  \Phi \right|^R_{-R},
\label{Ndot}
\ee
where we have decomposed $\psi$ as $|\psi| e^{-i \theta}$
and defined 
\be
\Phi(x)= i ( \psi_x \psi^* -\psi_x^* \psi).
\label{Phi}
\ee
The first term in the right-hand side of \eqref{Ndot} gives the rate at which the energy is pumped into the soliton
while 
the second one quantifies the damping rate.
Assuming that the interval $(-R,R)$ is large enough to contain 
the core of the soliton,  the last term in \eqref{Phi}  measures the radiation 
flux  through its endpoints.

The radiation through the points $x=\pm R$ outside the 
soliton's core is governed by the linearisation of \eqref{NLS}
whose dispersion relation is
 \be 
(\omega -i \gamma)^2= (k^2+1)^2-h^2.
\label{dr}
\ee
The radiation waves emanate from the oscillating soliton
which plays the role of a pacemaker for these waves.
The pacemaker has a 
 a period
$T$ and will sustain waves with 
 frequencies
$\omega= 2 \pi n /T$ ($n=1,2,...$); hence $\omega$ should be taken real in \eqref{dr}.
For real $\omega$,  equation \eqref{dr} has two pairs of complex roots
$k= \pm (p_{1,2} + iq_{1,2})$, with the imaginary components
\be
q_{1,2}= \sqrt{
\frac{1}{2} \pm \frac{{\cal S}}{2}
+ \frac12 \sqrt{1+2({\cal Q} \pm  {\cal S})}}, \label{q12}
\ee
where
\[
{\cal S}=\sqrt{{\cal P}+{\cal Q}},  \quad   {\cal Q}=\sqrt{{\cal P}^2+ \gamma^2 \omega^2},
\]
and $2{\cal P}=h^2-\gamma^2 +\omega^2$. 

The decay rate $q_1$ is always greater than $q_2$
and bounded from below ($q_1 > 1$);  it accounts for rapidly decaying core of the soliton.
Outside the core,
   the decay rate crosses over
to $q_2$, and if $q_2$ is small the solution enters the long oscillatory wing
 decaying in proportion to $e^ {-q_2 |x|}$.
Periodic solutions that we have reported in this paper
all have $\omega \sim 1$; for these,  the exponent $q_2$ can be small only if 
$\gamma$ is small.

Assuming that  $\gamma$ is small, the expression \eqref{q12} simplifies. Defining  the quantity 
\be
\sigma=\sigma(\omega)= \sqrt{ \omega^2+h^2-\gamma^2}-1,
\label{sigma}
\ee
we have, in particular, 
\begin{subequations} \label{sigamma} 
\begin{align}
q_2   \to 
\frac{\omega/2}{1+\sigma}  \left(\frac{\gamma}{\sigma}\right)^{1/2} \gamma^{1/2}
\quad   & \text{as}   \ \frac{\gamma}{\sigma} \to 0
\  & (\sigma>0);
\label{q2} \\
q_2  \to    \sqrt{\frac{\omega/2}{1+\sigma}}  \, \gamma^{1/2} \quad  & \text{as}  \ \frac{\sigma}{\gamma} \to 0; \\
q_2   \to  (-\sigma)^{1/2} \quad &  \text{as}  \ \frac{\gamma}{\sigma} \to 0 \  & (\sigma<0).
\end{align}
\end{subequations}
Thus $q_2$ is small if $\gamma$ is small 
while $\sigma$ is  positive. In particular, if $\sigma$ is much greater than $\gamma$, the exponent
$q_2 \sim \gamma/\sqrt{\sigma}$ while if $\sigma$ is much smaller than $\gamma$, we have
 $q_2 \sim  \sqrt{\gamma}$. The decay rate stays small ($\sim \sqrt{-\sigma}$) when $\sigma$ becomes
 negative --- as long as it remains  much smaller than $\gamma$
 in absolute value. However, as $\sigma$ grows to larger negative values,
 the exponent $q_2$ grows as well. 
Hence the quantity $\sigma(\omega)$ serves as an indicator of whether the 
long-range radiation with the frequency $\omega$ can be excited (i.e. whether $q_2$ is small)
or not.


\subsection{Radiation frequency selection}

The harmonic-wave solution of the linearised equation with the decay rate $q_2$ is 
\be
\psi = \eta \left( \cos \phi +
i \frac{\alpha \sin \phi- \beta \cos \phi}{({\cal S}+h)^2+ {\cal T}^2}
\right) e^{q_2x},
\label{hw}
\ee
where the phase $\phi=\omega t- p_2x$, the wavenumber $p_2=-{\cal T} /(2q_2)$, and
the coefficients
\[
\alpha= \omega ({\cal S}+h) + \gamma {\cal T}, \quad
\beta= \gamma({\cal S}+h) -  \omega {\cal T},
\]
and ${\cal T}=\sqrt{{\cal Q}-{\cal P}}$. The amplitude $\eta=\eta(\omega)$ is 
arbitrary as far as  the linearised equation is concerned but acquires a specific
value when \eqref{hw} is used to represent radiation from the soliton.

If $R$ is large enough,  the solution $\psi(x,t)$ with $x$ near $\pm R$ is described
by \eqref{hw}.
Substituting \eqref{hw} in \eqref{Phi} we obtain the flux
through the points $x=\pm R$:
\be
\Phi(\pm R)= \mp |\eta|^2  \frac{\cal T}{q_2}  \, \frac{ \omega({\cal S}+h) +\gamma {\cal T} }
{({\cal S}+h)^2+{\cal T}^2} \, e^{-2q_2R}.
\label{flux}
\ee

Assume, first, that $q_2$ is of order $1$;  this occurs, in particular, when $\gamma \sim 1$.
Since $R$ is chosen to be large,  the  flux through the points $x=\pm R$ in this case 
is exponentially small:
strongly damped radiation waves decay so quickly that they do not reach beyond the
core of the soliton (and essentially form part of it). 
The exponential factor in \eqref{flux} is not exponentially small only if 
$q_2$ is small; this, in turn, may only happen when the damping is weak.

These considerations have a simple physical interpretation.
When $\gamma$ is large, the soliton dissipates most of its excess energy within its core;
this is accounted for by the second term in \eqref{Ndot}.  On the other hand,
when $\gamma$ is small, the dissipation within the core
is insufficient to balance the first term in \eqref{Ndot}.
The long-range radiation may provide an alternative energy-draining mechanism in this case,
but this is  only available when $q_2$ is small which,
in turn, is indicated by $\sigma>0$.

If the period of the solution is  small enough for the quantity $\sigma(2 \pi/T)$ to be positive
and much greater than $\gamma$
(which is assumed small), the decay rate $q_2$ is close to zero. Therefore, the flux
\eqref{flux} is not exponentially small and the first-harmonic  radiation 
is not blocked in this case.
As $T$ is increased so that  $h^2-\gamma^2+ (2 \pi/T)^2$ drops below 1, 
the decay rate $q_2$ grows to values of order $1$. According to eq.\eqref{flux},
the 
first-harmonic radiation will be mostly suppressed in this case (i.e. damped outside a small neighbourhood of the soliton's core).
However, $\sigma(\omega)$ with $\omega= 4 \pi/T$ will be still positive; this  means that the tails will be 
dominated by the second-harmonic radiation now. Once the second harmonic is suppressed,
the third harmonic will take over, and so on.

We note that the second-harmonic radiation is present not only when the first-harmonic
radiation is suppressed. However, the amplitude $\eta(2 \pi n/T)$ typically scales
as $\epsilon^n$ and so Eq.\eqref{flux} implies that the second-harmonic
flux is much weaker than the first-harmonic flux in situations where both channels are open.

\subsection{First-harmonic radiation as a stabilising agent}

We routinely calculated the indicator $\sigma(2 \pi /T)$ as we 
continued our periodic solitons. The frequency spectrum of the radiation 
shows a remarkable correlation with the type of transition the soliton
is suffering.

The indicator \eqref{sigma}  with $\omega= 2 \pi/T$ is 
{\it positive\/} along each curve in Fig.\ref{ET_03} ---
moreover, $\sigma(\omega)$ is much greater than $\gamma$.
Therefore, 
in the case $\gamma>0.275$, the long-range radiation from the
soliton is dominated by the first harmonic. On the other hand, 
$\sigma(2\pi/T)$ is {\it negative\/} along each
$T1$ branch  in Fig.\ref{weak_damping_fig} which implies that the 
first-harmonic radiation is short-range when  $\gamma$ is smaller than $0.25$.
(When $\gamma=0.10$, $\sigma$ decreases from values close to $-\gamma$ to about
$-6 \gamma$ along the curve;
when $\gamma=0.20$, $\sigma$ drops from zero to nearly $-2.5 \gamma$.)
In the  latter case ($\gamma=0.10$ and $0.20$) the indicator $\sigma(\omega)$ becomes positive if
one lets $\omega= 4 \pi/T$; hence the radiation is second-harmonic
here (except for very large $T$ where it is third-harmonic). 

This observation allows us to understand, in qualitative terms, why the
transformation of the soliton with $\gamma>0.275$  is different from 
the scenario realised for very weak dampings ($\gamma<0.25$). When $\gamma$ is not very small, the soliton can
dispose the excess energy via two channels. Some energy will be lost
to the damping in the core region, where $|\psi| \sim 1$ and the term 
$i \gamma \psi$ is $\sim \gamma$. The rest will be sent away in the 
form of the strong first-harmonic radiation. 
Due to the availability of this powerful back-up dissipation channel,
the soliton behaves as an overdamped system; hence its enhanced stability.
On the other hand, when $\gamma$ is very small, the dominant long-range radiation is second harmonic.
Since the amplitude of the $n$-th harmonic
radiation scales as $\epsilon^n$ where $\epsilon$ is the scale of the amplitude of
the oscillation in the core, the second-harmonic radiation is much weaker than the
first-harmonic one, and the radiation channel is effectively blocked. The soliton shows the volatility
of an underdamped system; hence its instability and period doublings.

The formation of three-soliton complexes  observed 
in the continuation of the oscillatory solitons with $\gamma=0.30$ and $\gamma=0.35$  is also made
possible due to the availability of strong radiation with the period equal to the 
period of the central soliton. The lateral humps in the complex are seeded by the two maxima of the 
radiation sinusoid which are closest to the  central soliton.
On the other hand, the second-harmonic radiation available in the 
$\gamma=0.10$ and $\gamma=0.20$ cases is unable to seed the lateral solitons
because  it is  weak and  has the period different from the period of the central hump.

It is interesting to test these qualitative considerations using the case
of the intermediate damping, $\gamma=0.265$. In this case the  first-harmonic indicator
$\sigma(2\pi/T)$ is positive but $\sigma/\gamma$ is {\it not\/} much greater than 1
over the most of  the top $T1$ branch in
Fig.\ref{Fig_265}.
(As $h$ grows from $0.35$ to $0.71$ to $0.78$, the quotient
$\sigma/\gamma$ on the top branch changes from $0.5$ to $0$  to $1.$).
Accordingly, the first harmonic is strongly damped
($q_2 \sim \sqrt{\gamma}$) and the long-range radiation is second harmonic.
As a result, the soliton suffers period-doubling bifurcations, as
in the case of $\gamma=0.10$ and $0.20$. However, as we continue
further,
the quotient $\sigma/\gamma$ grows (to $2.$ at $h=0.82$)
and the first-harmonic radiation takes over
from the second-harmonic. Consequently, the solution transforms into
a three-soliton complex --- as in the case of $\gamma=0.30$ and $0.35$).

\section{Discussion and conclusions}
\label{DC}

\subsection{Homoclinic explosion}
\label{explosion}

The numerical continuation of the T1 solution with $\gamma< 0.25$
shows an unbounded growth of its period. The same is true for the T2
solution with $\gamma< 0.275$. These observations imply the presence of a 
homoclinic bifurcation where the period becomes infinite. We now argue 
that this bifurcation is a source of chaos.

We use results of Shilnikov \cite{Shilnikov_1}
who considered a three-dimensional dynamical system with a homoclinic orbit 
connecting 
a saddle-focus  to itself. The saddle-focus is a fixed point with
 a real unstable eigenvalue $\lambda_1>0$
and  two complex-conjugate stable eigenvalues $\lambda_2=\lambda_3^*$, 
 ${\rm Re\/} \, \lambda_{2,3}<0$.
Assuming that 
\be
\lambda_1 > | {\rm Re\/} \, \lambda_{2,3}|,
\label{Shil}
\ee
Shilnikov proved that
at the bifurcation point,  the system has infinitely many Smale's horseshoes 
in a neighbourhood of the homoclinic orbit \cite{Shilnikov_1} (see also \cite{Shilnikov_2}). 
Each of these horseshoes contains an invariant Cantor set with 
a countable infinity of (unstable) periodic orbits and an uncountable infinity 
of (unstable) bounded aperiodic orbits --- the strange invariant set 
which manifests itself as attractive or transient chaos.
Later it was shown that finitely many of the horseshoes persist on one side of the homoclinic
bifurcation \cite{GG}.
The homoclinic bifurcation giving birth to this plethora of periodic and chaotic
orbits has been referred to as the 
 ``homoclinic explosion" \cite{Sparrow}.
 
 The homoclinic explosion may also occur in four- and higher-dimensional systems 
 with a homoclinic orbit.  The necessary conditions for this  are formulated
 in terms of the eigenvalues of the linearisation about the fixed point connected to itself by the orbit: (a) out of all
 eigenvalues in the right half of the complex plane, the closest to the imaginary axis 
is a real eigenvalue $\lambda_1$; (b)
 in the left half of the complex plane, a pair of complex-conjugate eigenvalues $\lambda_{2,3}$
 are the closest  eigenvalues to the imaginary axis.  The homoclinic explosion occurs then if 
 the Shilnikov inequality \eqref{Shil} is satisfied \cite{Shilnikov_2}.  A carefully studied and clearly explained
 example of  homoclinic explosion in  partial differential equations is in
 \cite{Knobloch}.

 The role of the saddle-focus fixed point in our equation \eqref{NLS} is played by the
 soliton $\psi_-$, Eq.\eqref{solmin}.
 The linear spectrum  of this fixed point consists of two 
 discrete eigenvalues, Eqs.\eqref{lambdas} of the Appendix,  and a continuum of values
 $\lambda= -\gamma+i \omega$, 
 where $|\omega| \ge \omega_0$, $\omega_0=\sqrt{1-h^2}$.
  In the left half of the complex
 plane, the eigenvalue $\lambda_2$ 
 is further away from the imaginary axis than the continuous spectrum; hence 
 the dynamics near the saddle-focus will be dominated by  the single positive eigenvalue $\lambda_1$
 and the continuum of eigenvalues with  negative real part. This is not exactly
 the situation of Shilnikov; however it is similar
 in the sense that the fixed point has a one-dimensional unstable manifold
 and a more-than-one-dimensional stable manifold, formed by orbits 
 spiralling into the focus. It is natural to expect
 Shilnikov's result to remain valid in this case, provided an
 analog of the inequality \eqref{Shil}
 is in place --- although we 
 do not have any rigorous proof of that. The analog of the Shilnikov inequality  in our case 
 is
  \be
  \lambda_1 > \gamma,
\label{inne}
\ee
with $\lambda_1$ as in \eqref{lambdas}.

 The inequality \eqref{inne} translates into $(A_-^2/ \sqrt{3}) \Lambda_0 > \gamma$,
where $\Lambda_0(\epsilon)$ is the positive eigenvalue of the problem
\eqref{La}.
One can easily demarcate the region where this inequality is valid.
From \eqref{epsie} we have
\be
h=\sqrt{\left( \frac{\epsilon}{2+\epsilon} \right)^2+ \gamma^2 }.
\label{heps}
\ee
Equation \eqref{heps} together with 
\be
\gamma =  \frac{ A_-^2 }{\sqrt{3}}  \Lambda_0(\epsilon)
\label{geps}
\ee
define the parametric curve $h=h_{\rm Sh}(\gamma)$ on the $(\gamma,h)$ plane. The curve is 
parabolic in shape (see the dotted curve in the main frame and inset in
fig.\ref{chart_b}); it connects the origin to the point
$\gamma=0$, $h=1$. The inequality \eqref{inne} is valid in the region bounded by this curve
and the $h$-axis.
This region obviously includes the loci of the homoclinic bifurcations for all $\gamma<0.275$
(see fig.\ref{chart_b}). Therefore the homoclinic bifurcation is indeed  of the 
homoclinic-explosion type in our case. It is this bifurcation that is responsible for the 
appearance of temporally chaotic solitons in the parametrically
driven damped nonlinear Schr\"odinger equation.

\subsection{Attractor charts}

1. Our analysis  answers a number of 
questions raised by the attractor chart 
\cite{Bondila} compiled via direct numerical simulations. 
One such question concerns the existence of two types of the
transformation scenario of the parametrically driven soliton,
as the driving strength is raised for the fixed damping coefficient.
Is the transformation followed by the soliton with large $\gamma>0.275$ really 
different from the transition to chaos suffered by its weakly damped counterpart
($\gamma<0.25$), or
is this difference caused merely by numerical approximations?

The answer to this question is 
provided by diagrams in Figs. \ref{ET_03}, \ref{weak_damping_fig},  and \ref{Fig_265}.
In the region $\gamma<0.25$ 
(represented by the values $\gamma=0.10$ and $0.20$,
Fig.\ref{weak_damping_fig}) the increase of $h$ results in
 a sequence of 
period-doubling bifurcations of the periodic soliton. On
the contrary, no period-doubling
bifurcations occur as $h$ is increased for the fixed $\gamma>0.275$.
(See Fig.\ref{ET_03} showing the diagrams for $\gamma=0.30$ and $0.35$).

2. Another question is why no periodic solitons
are observed in numerical simulations with sufficiently large $h$.
 Is there a well-defined
boundary of the domain of existence of stable 
periodic solitons? We have shown that for
$\gamma>0.275$, the region of existence of periodic solitons 
on the $h$ axis is bounded by the saddle-node bifurcation. 
No such solutions, stable or
unstable, exist above this turning point. 
For small $\gamma$, $\gamma<0.25$, the region of existence 
of periodic solitons is also bounded by a turning point;  however,
solitons cease to exist as attractors for lower values of $h$.
The boundary of the stability domain here is set by a
narrow strip of temporally-chaotic solitons.

These conclusions verify and explain the formation of the ``desert region" on the
$(\gamma,h)$ plane. The positions of the saddle-node bifurcation point
for $\gamma >0.275$ and the accumulation point of higher-periodic solutions
for $\gamma<0.25$, are in good agreement with the boundary of the 
desert region detected in simulations \cite{Bondila}.

3. The third issue clarified by our analysis pertains to the
shape of the
region on the ($\gamma,h$)-plane where periodic solitons are observed in
numerical simulations. The question here is why does the $h(\gamma)$ 
curve bounding this region fold on itself for $\gamma$ 
between $0.25$ and $0.275$. 

This phenomenon is accounted
for by the restabilisation of the periodic soliton as $h$ is increased
for $\gamma=0.265$  (Fig.\ref{Fig_265}).
We note that the case $\gamma=0.265$ is intermediate between
the small- and large-$\gamma$ transformation scenarios:
On the one hand, similar to the weakly-damped scenario,
the soliton  undergoes a period-doubling cascade
as $h$ is raised for this $\gamma$ (more precisely, it undergoes
{\it two\/} period-doubling cascades). On the other,  the domain occupied by
the periodic attractor  is bounded by a saddle-node bifurcation in this case, 
as in the strongly-damped situation.

\begin{figure}
\includegraphics[width = \linewidth]{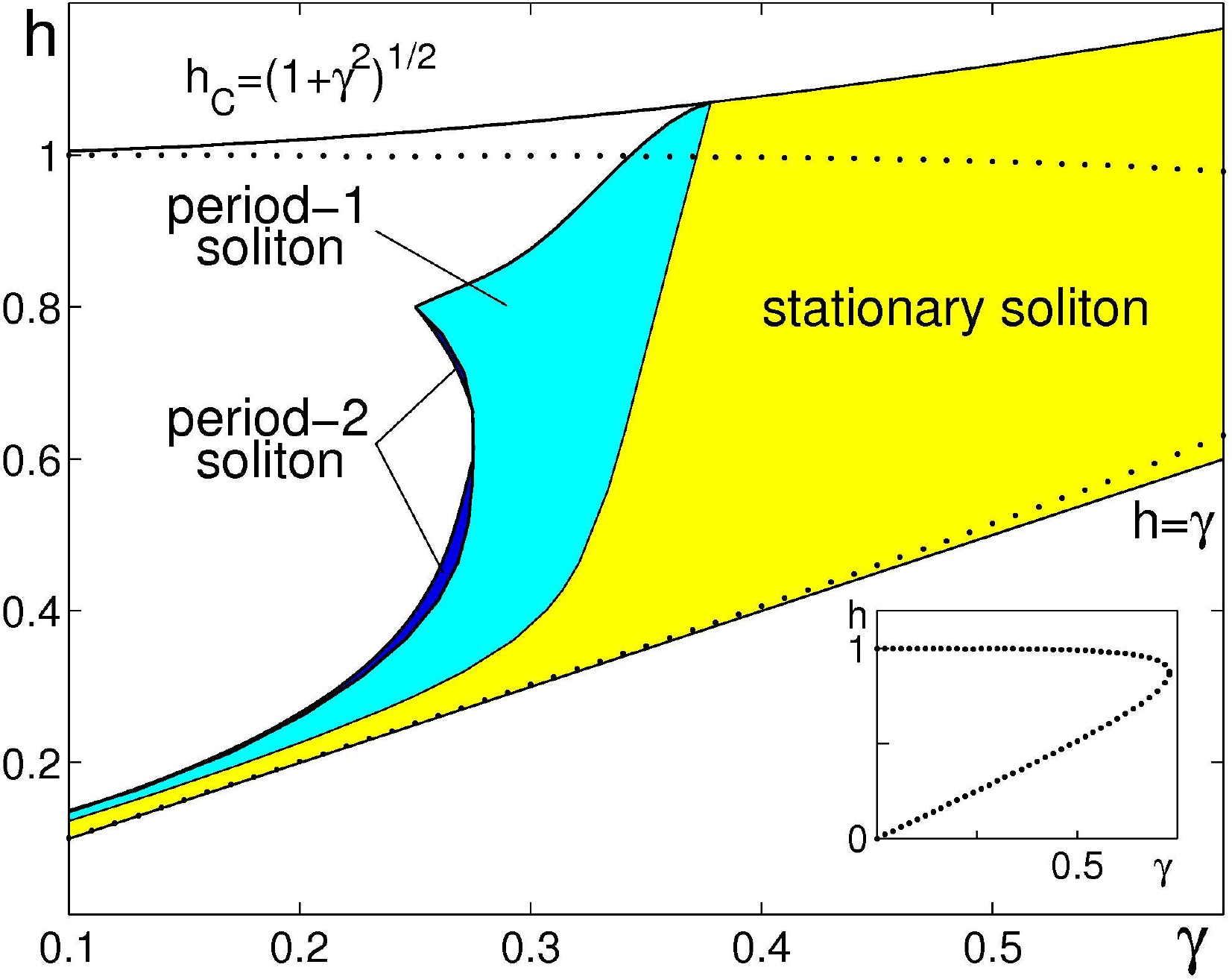}
\vspace*{1mm}
\caption{\label{chart_b}
(Color online) 
 Single-soliton attractor
chart.
The Shilnikov inequality \eqref{inne} holds
inside the region bounded by the dotted curve. 
The inset gives a wider perspective of this region.
}
\end{figure}

Fig.\ref{chart_b} summarises our conclusions on single-soliton periodic 
attractors. This diagram is in a good agreement with the attractor chart produced using  direct numerical
simulations, Fig.\ref{chart_from_PhysicaD}.

\appendix
\section{Eigenvalues of the soliton $\psi_-$}
\label{A}

We let $\psi=\psi_-(x)+\delta \psi(x,t)$, 
 where
  \[
 \delta \psi= e^{-i \theta_-}[u(x,t)+iv(x,t)]
 \]
  is a small perturbation, and
 linearise Eq.\eqref{NLS} in $\delta \psi$. Assuming that 
 \[
 u(x,t)= {\rm Re\/}  \left[ e^{\lambda t} p(x) \right],
 \quad
 v(x,t)= {\rm Re\/}  \left[ e^{\lambda t} q(x) \right]
 \]
 with complex $p$, $q$ and $\lambda$, yields an eigenvalue problem
 \be
 L_1 p+ 2 {\tilde \gamma} q = - {\tilde \lambda} q, \quad
 (L_0+ \epsilon) q = {\tilde \lambda} p.
 \label{EV_ori}
 \ee
Here we have defined ${\tilde \gamma}= \gamma/A_-^2$ and ${\tilde \lambda}= \lambda/A_-^2$, and 
introduced two Sturm-Liouville operators with familiar spectral properties:
\begin{align*}
L_0=-d^2/dX^2+1-2 \, {\rm sech}^2 X, 
\\
 L_1=-d^2/dX^2+1-6 \, {\rm sech}^2 X.
 \end{align*}
The independent variable $X= A_-x$ and the parameter $\epsilon$ was defined by
\be
\epsilon= 2 \sqrt{h^2-\gamma^2} /A_-^2.
\label{epsie}
\ee

The two-parameter 
 eigenvalue problem \eqref{EV_ori} can be reduced \cite{BBK} to a one-parameter problem 
 \be
 L_1 p= -\Lambda {\tilde q}, 
 \quad
 (L_0+ \epsilon) {\tilde q}= \Lambda p
 \label{La}
 \ee by letting 
 \be
 \Lambda^2=\lambda (\lambda + 2 {\tilde \gamma})
 \label{trsf}
 \ee
  and 
 ${\tilde q}= (\lambda+ 2 {\tilde \gamma}) \lambda^{-1} q$.

The lowest eigenvalue of the operator $L_0$ is zero; hence for any
$\epsilon>0$ the operator $L_0+\epsilon$ is positive definite  and the 
eigenvalue problem \eqref{La} can be cast in the form
\be
L_1 p= -\Lambda^2 M p,
\label{LM} 
\ee
where $M=(L_0+\epsilon)^{-1}$.
The operator in the left-hand side is symmetric and the 
one on the right is symmetric and positive definite; hence
all eigenvalues $-\Lambda^2$ are real and the function $p(x)$ can also be considered real.
The smallest eigenvalue of \eqref{LM} can be found as a minimum of the Rayleigh quotient:
\be
-\Lambda^2_0= \min_{p \in {\cal L}^2} \frac{(p, L_1 p)}{(p,Mp)},
\label{Ray0}
\ee
where $(p,q)$ stands for the ${\cal L}^2$ scalar product: $(p,q)= \int p(x) q(x)dx$.

The operator $L_1$ has a negative eigenvalue, 
$\eta_0=-3$, with the eigenfunction
$y_0= {\rm sech}^2 X$. Letting $p=y_0$ in Eq.\eqref{Ray0}, we conclude that 
the minimum is negative and hence
the problem \eqref{LM} has a negative eigenvalue.
 We denote the associated eigenfunction $p_0(x)$: 
\[
L_1 p_0= -\Lambda_0^2 M p_0.
\]
Since $L_1$ does not have any other negative eigenvalues
(the only other discrete eigenvalue of $L_1$ is $\eta_1=0$),
the problem  \eqref{LM} does
not have any other negative eigenvalues either. Indeed, 
assume there exist $-\Lambda_1^2$ and $p_1(x)$ such that 
$L_1 p_1= -\Lambda_1^2 M p_1$. The quadratic form $(y, L_1y)$ 
is then negative definite on the subspace spanned by $p_0$ and $p_1$;
in particular, it is negative for
$y(x)=C_0 p_0(x)+C_1 p_1(x)$ 
where the coefficients are chosen such that 
$(y,y_0)=0$. This, however, contradicts the fact that the form $(y,L_1y)$ 
cannot take negative values on the subspace orthogonal to $y_0$.

Thus the problem \eqref{La} has only one pair of nonzero eigenvalues,
 $\Lambda_0$ and  $-\Lambda_0$ (where $\Lambda_0$ is taken to be positive).
The function $\Lambda_0(\epsilon)$  has asymptotic  behaviours  $\Lambda_0 \to (4 \epsilon)^{1/2}$ as $\epsilon \to 0$
and  $\Lambda_0 \to (3 \epsilon)^{1/2}$ as  $\epsilon \to \infty$;
for intermediate values of $\epsilon$ we have tabulated it numerically.
Using eq.\eqref{trsf} we recover two eigenvalues of 
the original, two-parameter, problem \eqref{EV_ori}:
\be
\lambda_{1,2} = - \gamma  \pm 
\sqrt{\gamma^2+ A_-^4 \Lambda_0^2}.
\label{lambdas}
\ee
Here $\lambda_1>0$ and $\lambda_2< -\gamma<0$.

\acknowledgments

Remarks by Yuri Gaididei, Edgar Knobloch, Taras Lakoba and Bj\"orn Sandstede are  gratefully acknowledged.
IB was supported by the NRF of South Africa
(grants UID 65498, 68536 and 73608).
 EZ was supported by a DST grant under the
JINR/RSA Research Collaboration Programme and partially supported
by RFBR (grant No. 09-01-00770).


\begin{thebibliography}{99}


\bibitem{Miles}
J.W. Miles, J. Fluid Mech. {\bf 148}, 451 (1984)



\bibitem{LS} E W Laedke and K H Spatschek, J. Fluid Mech.{\bf 223}, 589 (1991)

%
\bibitem{Faraday}
C. Elphick and E. Meron, Phys. Rev. A {\bf 40}, 3226 (1989);
%
X.N. Chen and R.J. Wei, J. Fluid Mech. {\bf 259}, 291 (1994);
%
W. Zhang and J. Vi\~nals, Phys. Rev. Lett. {\bf 74}, 690 (1995);

A D D Craik and J G M Armitage, Fluid Dyn. Res. {\bf 15},  129 (1995); 
%
S P Decent and A D D Craik, J, Fluid Mech. {\bf 293} 237 (1995);
%
W. Wang, X. Wang, J. Wang, and R. Wei,
 Phys. Lett. A {\bf 219}, 74 (1996);
 %
 S P Decent, Fluid Dyn, Res. {\bf 21} 115 (1997);
%
X. Wang and R. Wei, Phys. Rev. Lett. {\bf 78}, 2744 (1997);
%
X. Wang and R. Wei, Phys. Rev. E {\bf 57}, 2405 (1998);
 %
G. Miao  and R. Wei,
  Phys.  Rev.  E {\bf 59}, 4075 (1999);
  %
W Chen, L Lu, Y Zhu, Phys. Rev. E {\bf 71}, 036622 (2005);
%
L Zhang, X Wang, and Z Tao, Phys. Rev. E {\bf 75}, 036602 (2007)
%


\bibitem{oscillons}
D. Astruc  and S. Fauve.  In: IUTAM Symposium on Free Surface
Flows. Proceedings of the IUTAM Symposium held in Birmingham, UK,
10-14 July 2000. A. C. King and  Y. D. Shikhmurzaev, editors.
Fluid Mechanics and Its Applications, Vol. {\bf 62} (Kluwer, 2001);
I V Barashenkov, N V Alexeeva, and E V Zemlyanaya, Phys. Rev.  Lett. {\bf 89}, 104101 (2002)

\bibitem{Clerc1}
M. G. Clerc, S. Coulibaly, and D. Laroze, Phys. Rev. E {\bf 77},   056209 (2008);
Int. J. of Bifurcation and Chaos, {\bf 19}, 2717 (2009);
{\it ibid.\/} 3525 (2009)
%
\bibitem{pendula} B. Denardo, B. Galvin, A. Greenfield, A. Larraza, S.
Putterman, and W. Wright,
 Phys. Rev. Lett. {\bf 68}, 1730 (1992);
%
W.-Z. Chen, Phys. Rev. {\bf B 49}, 15063  (1994);
%
G. Huang, S.-Y. Lou, and M. Velarde, Int. J. of Bifurcation and
Chaos, {\bf 6}, 1775 (1996);
%
N.V. Alexeeva, I.V. Barashenkov, and G.P. Tsironis, Phys. Rev.
Lett. {\bf 84}, 3053 (2000);
%
W. Chen, B. Hu, H. Zhang, Phys. Rev. {\bf B 65}, 134302 (2002);
%
H.-Q. Xu and Y. Tang, Chin. Phys. Lett. {\bf 23} 1544 (2006);
%
J. Cuevas, L. Q. English, P. G. Kevrekidis, and M. Anderson,
Phys. Rev. Lett. {\bf 102}, 224101 (2009)


\bibitem{optics} I.H.  Deutsch and I.  Abram,
J.  Opt.  Soc.  Am.  {\bf B 11}, 2303 (1994);
%
%
  A. Mecozzi, L. Kath, P.  Kumar, and C.G.  Goedde, Opt. Lett.  {\bf 19}, 2050 (1994);
%
 S. Longhi,  Opt. Lett.  20, 695 (1995);
 %
 V.J. S\'anchez-Morcillo, I. P\'erez-Arjona, F. Silva, G.J. de
Valc\'arcel, and E. Rold\'an, Opt. Lett. {\bf 25}, 957 (2000)

\bibitem{BBK} I.V.  Barashenkov, M.M.  Bogdan, and V.I.  Korobov,
 Europhys. Lett.  {\bf 15}, 113
(1991)

%
\bibitem{Clerc2} M. G. Clerc, S. Coulibaly, D. Laroze, Physica D {\bf 239}, 72 (2009);
Europhys. Lett. {\bf 90}, 38005 (2010)
%

\bibitem{JA} T.-C. Jo and D. Armbruster, Phys. Rev. E {\bf 68}, 016213 (2003).
See also the following references where the parametrically driven cubic Klein-Gordon 
equation is studied in the Fourier space: W. I. Newman, R. H. Rand, A. L. Newman,
Chaos, {\bf 9}, 242 (1999); T. Bakri, H. G. E. Meijer, F. Verhulst, J. Nonlinear Sci. {\bf 19}, 571 (2009)



%


%

\bibitem{BZ3} I V Barashenkov and E V Zemlyanaya, the next submission

\bibitem{Bondila} Bondila M, Barashenkov I V, Bogdan M M,
 Physica D {\bf 87} 314 (1995)

%
\bibitem{ABP} N V Alexeeva, I V Barashenkov, and D E Pelinovsky,
Nonlinearity {\bf 12} 103 (1999)
%
\bibitem{FLS} H Friedel, E W Laedke and K H Spatschek,
J. Fluid Mech. {\bf 284} 341 (1995)

%
\bibitem{continuation}
E.V. Zemlyanaya and I.V. Barashenkov,
 Math. Modelling  {\bf 16} 3 (2004)
%
\bibitem{Arnold}
V. I. Arnold, Mathematical Methods of Classical Mechanics
(Graduate Texts in Mathematics). Springer Verlag, New York (2010)



 \bibitem{AW}
D M Ambrose and J Wilkening, Commun. Appl. Math. Comput. Sci. {\bf 4}, 177  (2009); 
 J. Nonlinear Sci. {\bf 20}, 277 (2010)

\bibitem{Shilnikov_1} L P Shil'nikov, DAN SSSR {\bf 160}, 558 (1965)

\bibitem{Shilnikov_2} L P Shil'nikov, Mat. Sbornik {\bf 81} ({\bf 123}), 92 (1970)

\bibitem{GG} P. Glendinning and C. Sparrow, J. Stat. Phys. {\bf 35} 645 (1984)


\bibitem{Sparrow} C.  Sparrow, The Lorenz equations: bifurcations, chaos and strange attractors.
(Applied Mathematical Sciences {\bf 41}).
Springer-Verlag, New York, 1982


\bibitem{Knobloch} D. R. Moore, J. Toomre, E. Knobloch and N. O. Weiss. Nature {\bf 303}, 663 (1983);
E. Knobloch, D. R. Moore, J. Toomre, and N. O. Weiss, J. Fluid Mech. {\bf 166}, 409 (1986)

\end{thebibliography}
\end{document}